\documentclass[twocolumn,floatfix,superscriptaddress,amsmath,showpacs,showkeys,aps,prb]{revtex4}

\usepackage[final]{graphicx}
\usepackage{t1enc}
\usepackage{bm}

\begin{document}

\bibliographystyle{apsrev}

\title{\bf Ising nematic phase in ultra-thin magnetic films: a Monte Carlo study}

\author{Sergio A. Cannas}
\email{cannas@famaf.unc.edu.ar} \affiliation{Facultad de
Matem\'atica, Astronom\'{\i}a y F\'{\i}sica, Universidad Nacional
de C\'ordoba, \\ Ciudad Universitaria, 5000 C\'ordoba, Argentina}
\altaffiliation{Member of CONICET, Argentina}
\author{Mateus F. Michelon}
\email{michelon@ifi.unicamp.br} \affiliation{Departamento de F\'{\i}sica,
Universidade Federal do Rio Grande do Sul\\
CP 15051, 91501--979, Porto Alegre, Brazil}
\altaffiliation{Present address: Instituto de Física Gleb Whataghin, UNICAMP, Campinas, Brazil}
\author{Daniel A. Stariolo}
\email{stariolo@if.ufrgs.br}
\affiliation{Departamento de F\'{\i}sica,
Universidade Federal do Rio Grande do Sul\\
CP 15051, 91501--979, Porto Alegre, Brazil}
\altaffiliation{Research Associate of the Abdus Salam International Centre for Theoretical Physics,
Trieste, Italy}
\author{Francisco A. Tamarit}
\email{tamarit@famaf.unc.edu.ar} \affiliation{Facultad de  Matem\'atica, Astronom\'{\i}a
y F\'{\i}sica, Universidad Nacional de C\'ordoba, \\ Ciudad Universitaria, 5000
C\'ordoba, Argentina} \altaffiliation{Member of CONICET, Argentina}

\date{\today}

\begin{abstract}
We study the critical properties of a two--dimensional Ising model
with competing ferromagnetic exchange and dipolar interactions,
which models an ultra-thin magnetic film with high out--of--plane
anisotropy in the monolayer limit. We present numerical evidence
showing that two different scenarios  appear in the model for
different values of the exchange to dipolar intensities ratio,
namely, a single first order stripe - tetragonal phase transition
or two phase transitions at different temperatures with an
intermediate Ising nematic phase between the stripe and the
tetragonal ones. Our results are very similar to those predicted
by Abanov {\it et al} [Phys. Rev. B {\bf 51}, 1023 (1995)], but
 suggest a much more complex critical behavior than
the predicted by those authors for both the stripe-nematic and the
nematic-tetragonal phase transitions.

We also show that the presence of diverging free energy barriers
at the stripe-nematic transition makes possible to obtain by slow
cooling a metastable supercooled nematic state down to
temperatures well below the transition one.

\end{abstract}

\pacs{75.70.Kw,75.40.Mg,75.40.Cx}
\keywords{ultra-thin magnetic
films, Ising model}

\maketitle

\section{Introduction}

The study of thin magnetic films have deserved an increasing
interest during the last decade, both on their experimental and
theoretical aspects. Besides the great amount of technological
applications related to their magnetic behavior, as for instance,
data storage, these studies have also faced statistical physicists
with the challenge of trying to answer many foundational question
regarding the role of microscopic competing interactions on the
macroscopic static and dynamical behavior of two dimensional
systems. Moreover, the constant development of novel methods for
constructing ultra thin films, together with a significant
improvement in the quality of techniques for measuring nanoscopic
magnetic structures,e have open many fascinating questions, many
of them still open.

Many ultrathin magnetic films, like e.g. Co/Cu, Co/Au, Fe/Cu,
undergo a reorientation transition at a temperature $T_R$ in which
the spins align preferentially in a direction perpendicular to the
film ~\cite{AlStBi1990,VaStMaPiPoPe2000,DeMaWh2000}. This
reorientation transition is due to the competition between the
in-plane part of the dipolar interaction and the surface
anisotropy~\cite{Po1998}. Furthermore, in the range of
temperatures where the magnetization points out of the plane, the
competition between exchange and dipolar interaction causes the
global magnetization to be effectively zero but instead striped
magnetic domain patterns emerge\cite{BoMaWhDe1995,DeMaWh2000}. In
the limit when the stripe width is much larger than the domain
walls, the walls can be approximated by Ising walls and the system
can be considered as an Ising system of interacting domain
walls~\cite{Po1998}. In spite of intense
theoretical\cite{DeMaWh2000,BoMaWhDe1995,AbKaPoSa1995,SaAlMe1996,Po1998,StSi2000,ToTaCa1998,StCa1999,GlTaCa2002,GlTaCaMo2003,CaStTa2004,CaDaRaRe2004}
and
experimental\cite{AlStBi1990,VaStMaPiPoPe2000,PoVaPe2003,WuWoScDoZhJiQi2004,PoVaPe2006}
work on the behavior of ultrathin magnetic films, the precise
nature of the phases and the relaxational dynamics aspects of
these systems is still poorly understood. In Monte Carlo
simulations of an Ising model on a square lattice, Booth et
al.~\cite{BoMaWhDe1995} found evidence of a stripe phase a low
temperatures, with orientational and positional order reminiscent
of the smectic order in liquid crystals. These authors found a
transition from a stripe phase to a high temperature phase with
broken orientational order, called {\em tetragonal liquid} phase.
In this phase domains of stripes with mutually perpendicular
orientations emerge and form kind of labyrinthine patterns. At
still higher temperatures these domains collapse and the system
crosses over continuously to a completely disordered behavior.
 From numerical data for the specific heat these authors found
evidence of a continuous stripe/tetragonal-liquid transition.
However, it has been argued that the transition observed by Booth
et al should correspond to a nematic-tetragonal phase
transition\cite{Po1998}. Moreover, recent numerical results using
time series methods on the same model with the same parameters
values suggest that such transition is not continuous but it is a
first order one\cite{CaDaRaRe2004}. In an important contribution
Abanov et al.~\cite{AbKaPoSa1995} analyzed the different phases
which could emerge from a continuous approximation to the domain
wall crystal. They predicted two possible scenarios in zero
magnetic field. In the first scenario a smectic-like low
temperature phase with spatial correlations decaying algebraically
with distance appears at low temperatures.In this phase there is
{\em quasi long range order} (QLRO) and the orientational order
parameter is non zero. The proliferation of bound dislocations
pairs at higher temperature causes QLRO to be destroyed through a
Kosterlitz-Thouless phase transition to a nematic-like phase. In
this phase only orientational order is observed. At still higher
temperatures even orientational order is destroyed through the
appearance of domains of perpendicular stripes. This is the
tetragonal--liquid phase. Since $Z_2$ orientational symmetry is
restored at this transition the authors speculated that this
transition should be in the Ising universality class. In the
second scenario the system is not able to sustain a purely nematic
phase and goes from the smectic directly to the tetragonal liquid
through a first order transition. Recently Cannas et
al.~\cite{CaStTa2004} found numerical evidence of this second
scenario, also supported by a continuous approximation with a
Landau-Ginzburg Hamiltonian, from which a fluctuation-induced
first order transition is predicted. Up to now no evidence was
found of the first scenario predicted by Abanov et al. In this
paper we show that for large enough system sizes, Monte Carlo
simulations of the same model studied in
\cite{BoMaWhDe1995,CaStTa2004} support the appearance of a
sequence of two phase transitions of the kind predicted by Abanov
et al. when the width of the equilibrium stripes is large enough,
while for thin stripes the second scenario with a unique
transition is found.

We consider a system of magnetic dipoles on a square lattice in
which the magnetic moments
 are oriented perpendicular to the plane of the lattice, with both nearest-neighbor
ferromagnetic exchange interactions and long-range dipole-dipole
interactions between moments. The thermodynamics of this system is
ruled by the dimensionless Hamiltonian~\cite{difference}:
\begin{equation}
{\cal H}= - \delta \sum_{<i,j>} \sigma_i \sigma_j + \sum_{(i,j)}
\frac{\sigma_i \sigma_j}{r^3_{ij}} \label{Hamilton1}
\end{equation}
\noindent where $\delta$ stands for the ratio between the exchange
$J_0>0$ and the dipolar $J_d>0$ interactions parameters, i.e.,
$\delta = J_0/J_d$. The first sum runs over all pairs of nearest
neighbor spins and the second one over all distinct pairs of spins
of the lattice; $r_{ij}$ is the distance, measured in crystal
units, between sites $i$ and $j$. The energy is measured in units
of $J_d$. The overall (known) features of the equilibrium phase
diagram of this system can be found in
Refs.\cite{DeMaWh2000,MaWhRoDe1995,BoMaWhDe1995,GlTaCa2002,CaStTa2004,CaDaRaRe2004},
while several dynamical properties at low temperatures can be
found in Refs.\cite{SaAlMe1996,ToTaCa1998,StCa1999,GlTaCaMo2003}.
The threshold for the appearance of the stripe phase in this model
is $\delta_c=0.425$~\cite{MaWhRoDe1995,difference}. As $\delta$
increases the system presents a sequence of striped ground states,
characterized by a constant width $h$, whose value increases
exponentially with $\delta$ \cite{MaWhRoDe1995}.

We show that energy and orientational order parameter histograms
present a sequence of three peaks for $\delta=2$, corresponding to
three different thermodynamic phases, while for $\delta=1$ only
two peaks are observed, consistent with the presence of only one
phase transition. In the first case we show that the intermediate
phase presents long range orientational order but no long range
positional order,
 consistently with a nematic phase. Our results show that the low
 energy phase is a striped one. We did not find evidence of smectic order,
 although the possible existence of algebraic decaying correlations (near
 and below the transition temperature), strongly hidden by finite size effects,
 cannot be excluded. Finite size scaling analysis of specific heat and the fourth
order cumulant of the energy give evidence of a
 first order transition between the nematic
and tetragonal phases. However, the analysis of the orientational
order parameter histograms suggests the existence of more than one
nematic phases for much larger system sizes than the ones
considered here. On the other hand, the stripe-nematic phase
transition shows unusual features, some of them characteristic of
a first order transition, but some other properties strongly
resembling those observed in a Kosterlitz-Thouless (KT) phase
transition.

 For $\delta=1$ a unique weakly first order phase transition is
supported by direct thermodynamic analysis through a computation
of the free energy of the different phases. In the last section we
show that the previous thermodynamic behavior is also supported by
out of equilibrium measurements during quasi-static
cooling/heating cycles, where strong metastability is observed at
the stripe-nematic phase transition. This behavior is consistent
with the observation of asymptotically divergent free energy
barriers at this transition.

\section{The Ising nematic phase}
\label{histos}

 We first analyzed the equilibrium histograms $P(E/N)$
for the energy per spin at different temperatures $T$ and
different system sizes $L$. For every system size and every
temperature the corresponding energy histogram was calculated by
recording the energy values during a single MC run. Before
starting to record the energy we left the system run a transient
period of $M_1$ MC steps (MCS) in order to equilibrate. After that
period we recorded the energy values over $M_2$ MCS. A MCS is
defined as a complete cycle of $N$ spin update trials, according
to a heat bath dynamics algorithm. For every pair of values of $T$
and $L$ we checked different values of $M_1$ and $M_2$ in order to
ensure a stationary distribution $P(E/N)$. Typical values of $M_1$
were between $10^6$ and $2 \times 10^7$ MCS, while typical values
of $M_2$ were between $2 \times 10^7$ and $2 \times 10^8$ MCS. A
similar calculation was carried out to obtain the equilibrium
histograms of the orientational order parameter\cite{BoMaWhDe1995}

\begin{equation}
\eta \equiv \frac{n_v-n_h}{n_v+n_h} \label{eta}
\end{equation}

\noindent where $n_v$ ($n_h$) is the number of vertical
(horizontal) bonds between nearest neighbor anti--aligned spins.
This parameter takes the value $+1$ ($-1$) in a completely ordered
horizontal (vertical) striped state, while it equals zero in any
phase with $90^o$ rotational symmetry, thus describing the $90^o$
rotational symmetry breaking.

\begin{figure}
\begin{center}
\includegraphics[scale=0.5]{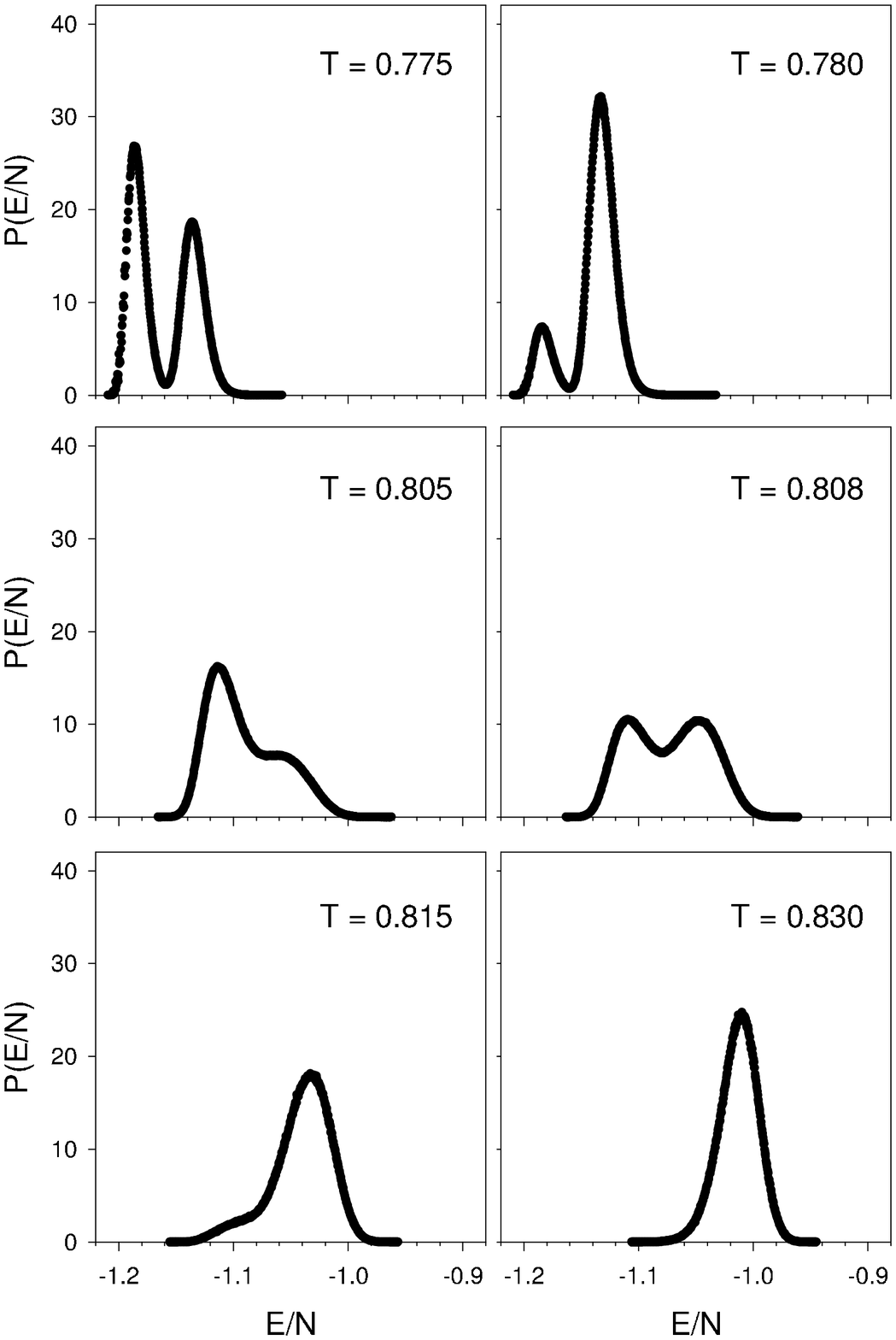}
\caption{\label{fig1} Energy per spin probability distributions
(normalized histograms) for $\delta=2$ and $L=56$.}
\end{center}
\end{figure}

\begin{figure}
\begin{center}
\includegraphics[scale=0.5]{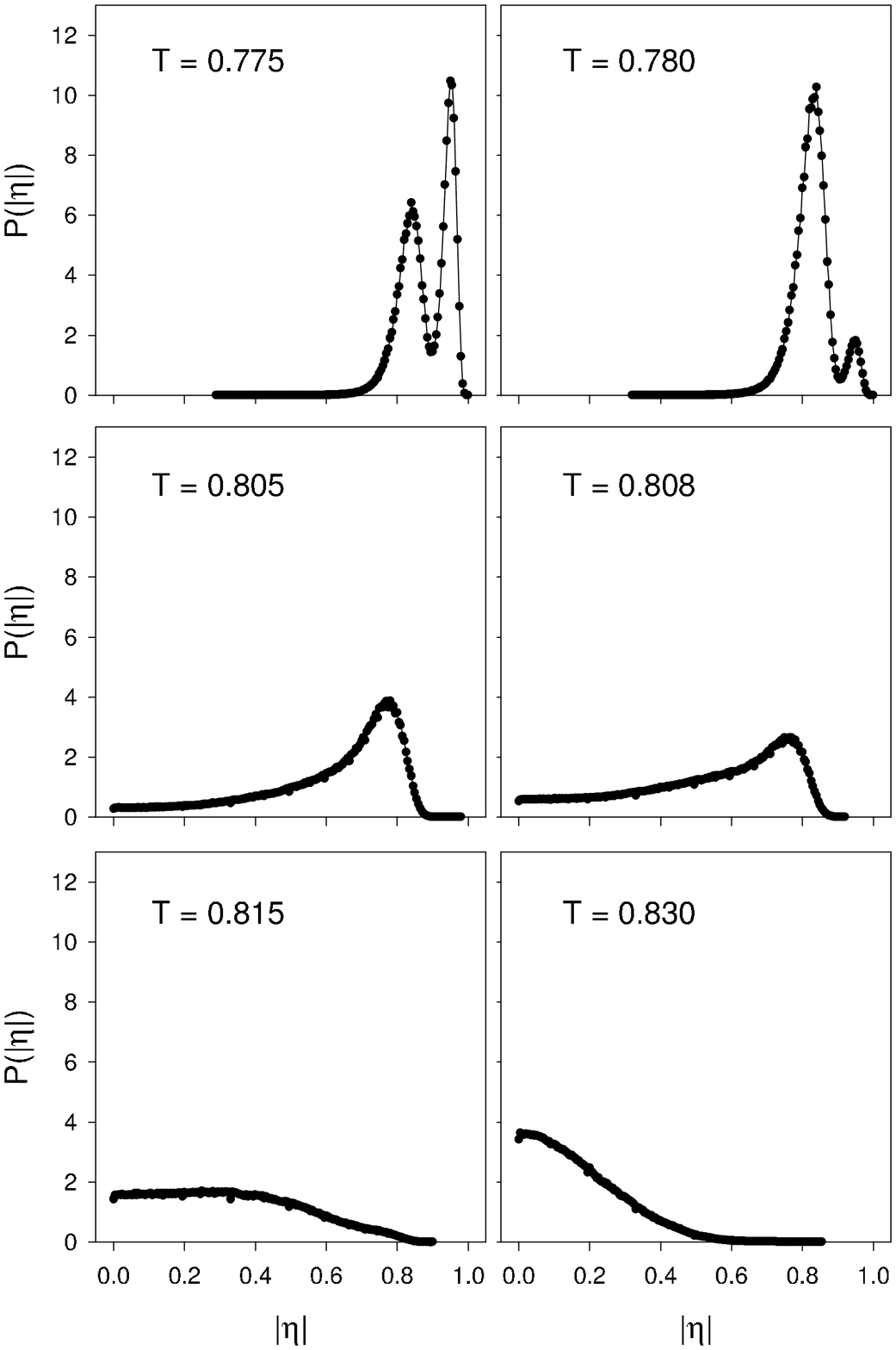}
\caption{\label{fig2} Orientational order parameter probability
distributions (normalized histograms) for $\delta=2$ and $L=56$.}
\end{center}
\end{figure}

We concentrated first in analyzing the $\delta=2$ case ($h=2$
striped ground state). In Fig.\ref{fig1} we show the behavior of
$P(E/N)$ for $L=56$, $\delta=2$ and different temperatures, while
in Fig.\ref{fig2} we show the corresponding histograms for the
distribution of the absolute value of the orientational order
parameter $P(|\eta|)$.

\begin{figure}
\begin{center}
\includegraphics[scale=0.3,angle=-90]{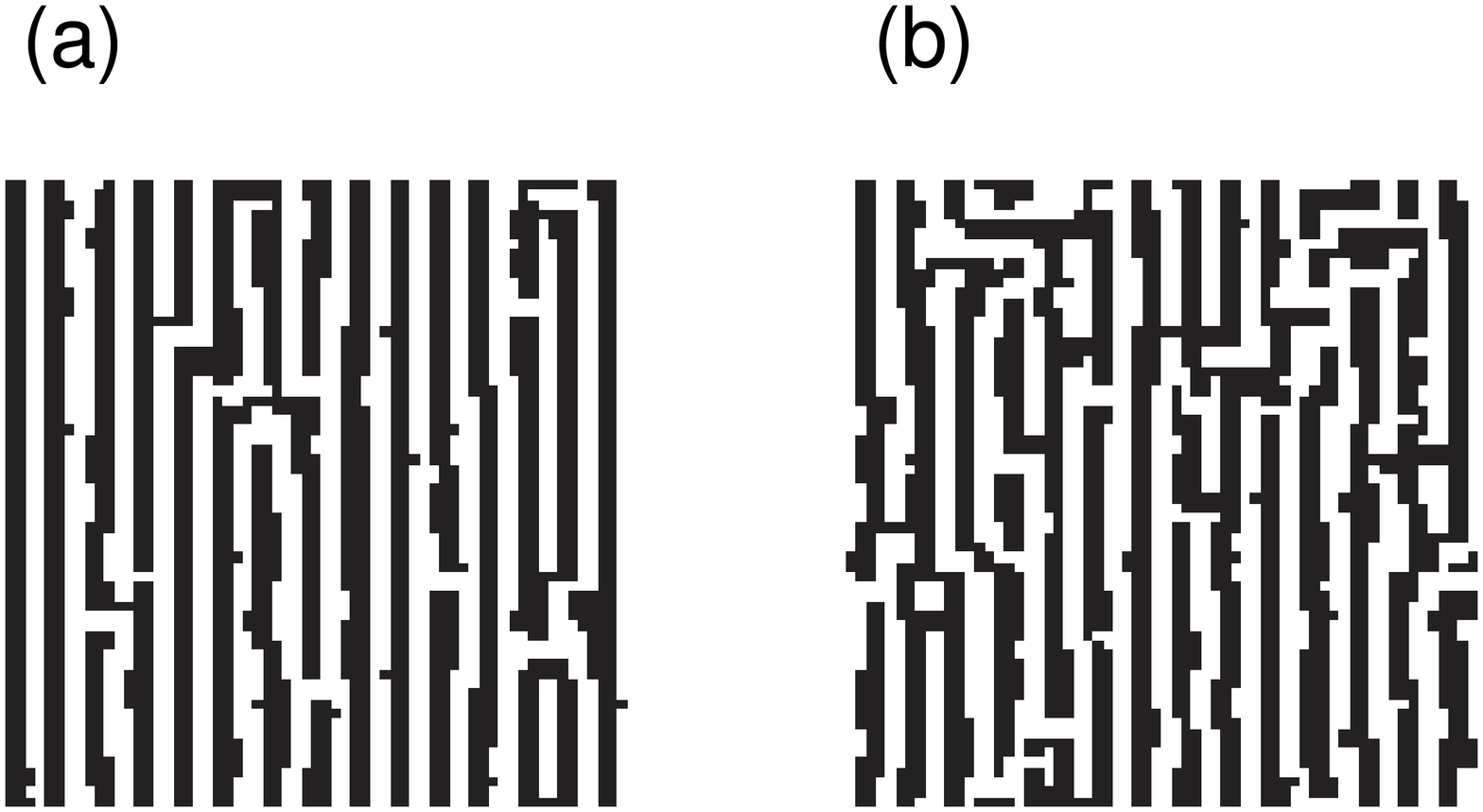}
\caption{\label{fig3} Typical spin configurations in the nematic
phases for $\delta=2$ and $L=64$. (a) $T=0.78$, the orientational
order parameter for this configuration is $|\eta | \approx 0.8$;
(b) $T=0.807$ and  $|\eta | \approx 0.59$}
\end{center}
\end{figure}

The presence of pairs of peaks in both the energy per spin and the
orientational order parameter distributions, located at three
distinct ranges of values, is a clear signature of the existence
of three different phases, with two phase transitions between
them. The low energy peak is associated with a peak in the order
parameter centered around $\left| \eta \right| \approx 1$, thus
corresponding to a phase with long range orientational order.
Moreover, an analysis of a stripe order parameter introduced below
(and also a direct inspection of the typical spin configurations
at those energies) shows that this is an ordered striped phase
with $h=2$. The highest energy peak is associated with a peak in
the order parameter centered at $\eta = 0$. A direct inspection of
the associated typical spin configurations at those energies shows
indeed that they correspond to a tetragonal liquid phase. On the
other hand, the intermediate energy phase is associated with a
peak of the order parameter centered around $\left| \eta \right|
\approx 0.81$, thus corresponding to a state with  broken $90^o$
rotational symmetry. In Fig.\ref{fig3}a we illustrate a typical
spin configuration in this intermediate phase for $L=64$ and
$T=0.78$. We see that this phase is characterized by a high
density of topological defects, mainly dislocations in the
directions of the underlying striped structure. The same type of
defects have been observed in Fe on Cu ultrathin films near the
phase transition (see Fig.2 in Ref.\cite{VaStMaPiPoPe2000}). Such
defects reduce the average value of the orientational order
parameter and their presence is in agreement with the qualitative
description of the Ising nematic phase introduced by Abanov and
coauthors \cite{AbKaPoSa1995}. To confirm this assumption we
calculated the spatial correlations along the coordinate
directions given by

\begin{eqnarray}
C_x(r) &\equiv& \frac{1}{N} \sum_y \sum_{x} \left<\sigma_{x,y}\;
\sigma_{x+r,y} \right> \\
C_y(r) &\equiv& \frac{1}{N} \sum_y \sum_{x} \left<\sigma_{x,y}\;
\sigma_{x,y+r} \right>
\end{eqnarray}

\begin{figure}
\begin{center}
\includegraphics[scale=0.38,angle=-90]{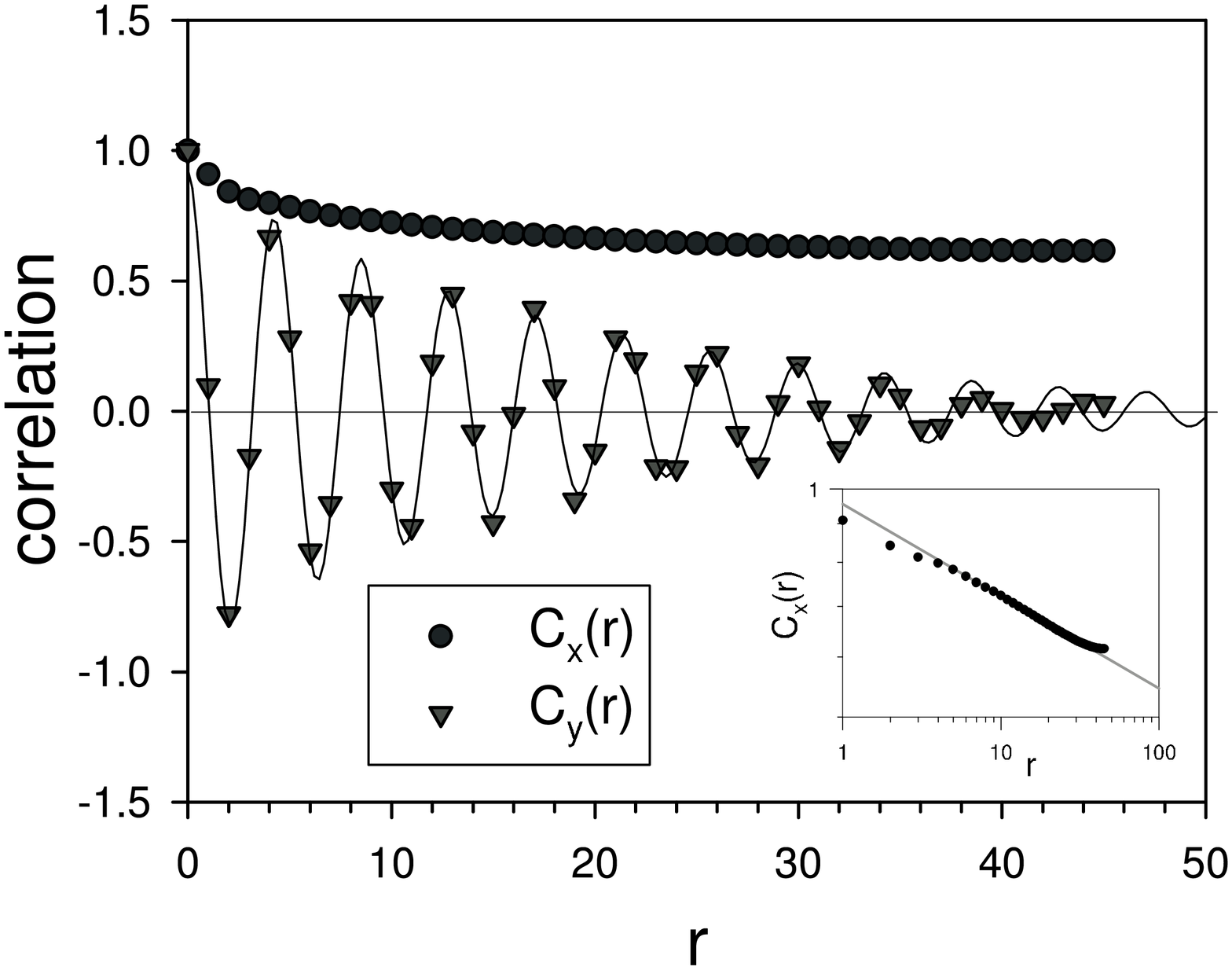}
\caption{\label{fig4} Spatial correlation functions along the
coordinate directions in a nematic phase for $\delta=2$, $T=0.78$
and $L=92$. The continuous line corresponds to a fitting using a
function $f(r)= A \exp{(r/\xi)\, \sin{(k_0\, r + \phi)}}$, with
fitting parameters $k_0 = 1.47$, $\xi = 17.4$ and $\phi=1.63$.
Notice that $k_0\approx \pi/2$, the wave vector of the striped
structure with $h=2$ (ground state for $\delta=2$). The log-log
plot of $C_x$ in the inset shows that it decays at large distances
with a power law $r^{-\omega}$, with an exponent $\omega \approx
0.12$.}
\end{center}
\end{figure}

An example of the behavior of $C_x(r)$ and $C_y(r)$ in the
intermediate phase is shown in Fig.\ref{fig4} for $L=92$ and
$T=0.78$. This figure shows clearly that the correlations decay
algebraically in one of the coordinate directions and
exponentially in the other.We see that the intermediate phase is
characterized by long range orientational order but does no t
present positional order. Although the description level of the
elastic approximation of Abanov and coauthors does not allow a
simple derivation of the spin-spin spatial correlations, the above
features are in qualitative agreement with their characterization
of the nematic phase.

To further characterize the different phases we defined a stripe
order parameter through the static structure factor

\begin{equation}
S(\vec{k})  \equiv \left< \left| \hat{\sigma}_{\vec{k}} \right|^2
\right>=\frac{1}{N^2} \sum_{\vec{r}} \sum_{\vec{r}\,
'}C(\vec{r},\vec{r}\, ') e^{i\vec{k}.(\vec{r}-\vec{r}\, ')}
\label{factor}
\end{equation}

\noindent where $\hat{\sigma}_{\vec{k}} = \frac{1}{N}
\sum_{\vec{r}} e^{i \vec{k}\vec{r}} \sigma_{\vec{r}}$ is the
discrete Fourier transform of $\sigma_{\vec{r}}$. It can be shown
that in a pure striped state of width $h$ the only non-zero
components correspond to $(k_x,k_y)=(0,\pm \pi p/h)$ (horizontal
stripes) or $(k_x,k_y)=(\pm \pi p/h,0)$ (vertical stripes), where
$p$ takes all the odd integer values between one and $h$. For
instance, in a pure $h=2$ vertical striped state  it can be shown
that

\begin{equation}
S(\vec{k}) =\frac{1}{2}\; \delta_{k_y,0}\;  \left(
\delta_{k_x,\pi/2}+  \delta_{k_x,-\pi/2}\right) \label{Sstripes}
\end{equation}

Now, since all the phases we are dealing with in this case present
a discrete rotational symmetry, either of $90^o$ or $180^o$
respect to the coordinate axes, the maxima of $S(\vec{k})$ will be
located at $(k_x,k_y)=(\pm k_0,0)$ and/or $(k_x,k_y)=(0,\pm k_0)$,
with $k_0$ some value close to $\pi/2$. Then we can define the
stripe order parameter as

\begin{equation}
\eta_s = 2 \left(S(k_0,0)+S(0,k_0) \right)
\end{equation}

\noindent (this definition can be easily generalized to consider
underlying ground states of arbitrary width $h$). This order
parameter will take values close to one for states with long range
$h=2$ stripe order and zero (in the thermodynamic limit) for any
state without long range positional order. For instance, let us
consider a system in which correlations decay exponentially in the
$x$ direction and remain almost constant in the $y$ direction,
namely, $C(\vec{r},\vec{r}\, ')=\exp{(-\lambda |x-x'|)}\;
\cos{(k_0\, x)}$; this is a rough approximation of the behavior we
observe in Fig.\ref{fig4}, where the algebraic decay in one of the
coordinate directions is extremely slow. Substituting into the
definition of $S(\vec{k})$ and replacing the sums by integrals, in
the large $L$ limit we get

\begin{eqnarray}
S(\vec{k}) &\sim& \frac{\delta_{k_y,0}}{L} \int_0^\infty
e^{-\lambda\, x} cos(k_0\, x) \cos{k_x\,x} \; dx \nonumber \\
\mbox{ } &\sim& \frac{\delta_{k_y,0}}{2L} \left(
\frac{\lambda}{(k_x-k_0)^2+\lambda^2}+
\frac{\lambda}{(k_x+k_0)^2+\lambda^2}\right)
 \label{Snematic}
\end{eqnarray}

\noindent If the correlation length $\xi_s =\lambda^{-1}$ is
independent of the system size we have that the stripe order
parameter goes to zero as $\eta_s \sim L^{-1}$ when
$L\rightarrow\infty$. In a tetragonal liquid state we can assume
for the correlation the following form

\begin{equation}
C(\vec{r},\vec{r}\, ') = e^{-\lambda (|x-x'|+|y-y'|)}\;
cos(k_0\,|x-x'|)\; cos(k_0\, |y-y'|)
\end{equation}

\noindent which possesses the tetragonal symmetry. We have then

\begin{eqnarray}
S(\vec{k}) &\sim& \frac{1}{4L^2} \left(
\frac{\lambda}{(k_x-k_0)^2+\lambda^2}+
\frac{\lambda}{(k_x+k_0)^2+\lambda^2} \right. \nonumber \\
\mbox{ } &+& \left. \frac{\lambda}{(k_y-k_0)^2+\lambda^2}+
\frac{\lambda}{(k_y+k_0)^2+\lambda^2} \right)
 \label{Stetragonal}
\end{eqnarray}

\noindent which reproduces the observed crown shape of the
structure factor observed in a tetragonal liquid state, both in
numerical simulations\cite{DeMaWh2000} and in Fe/Cu films
images\cite{PoVaPe2003}. In this case we have that the stripe
order parameter goes to zero as $\eta_s \sim L^{-2}$ when
$L\rightarrow\infty$.

\begin{figure}
\begin{center}
\includegraphics[scale=0.42]{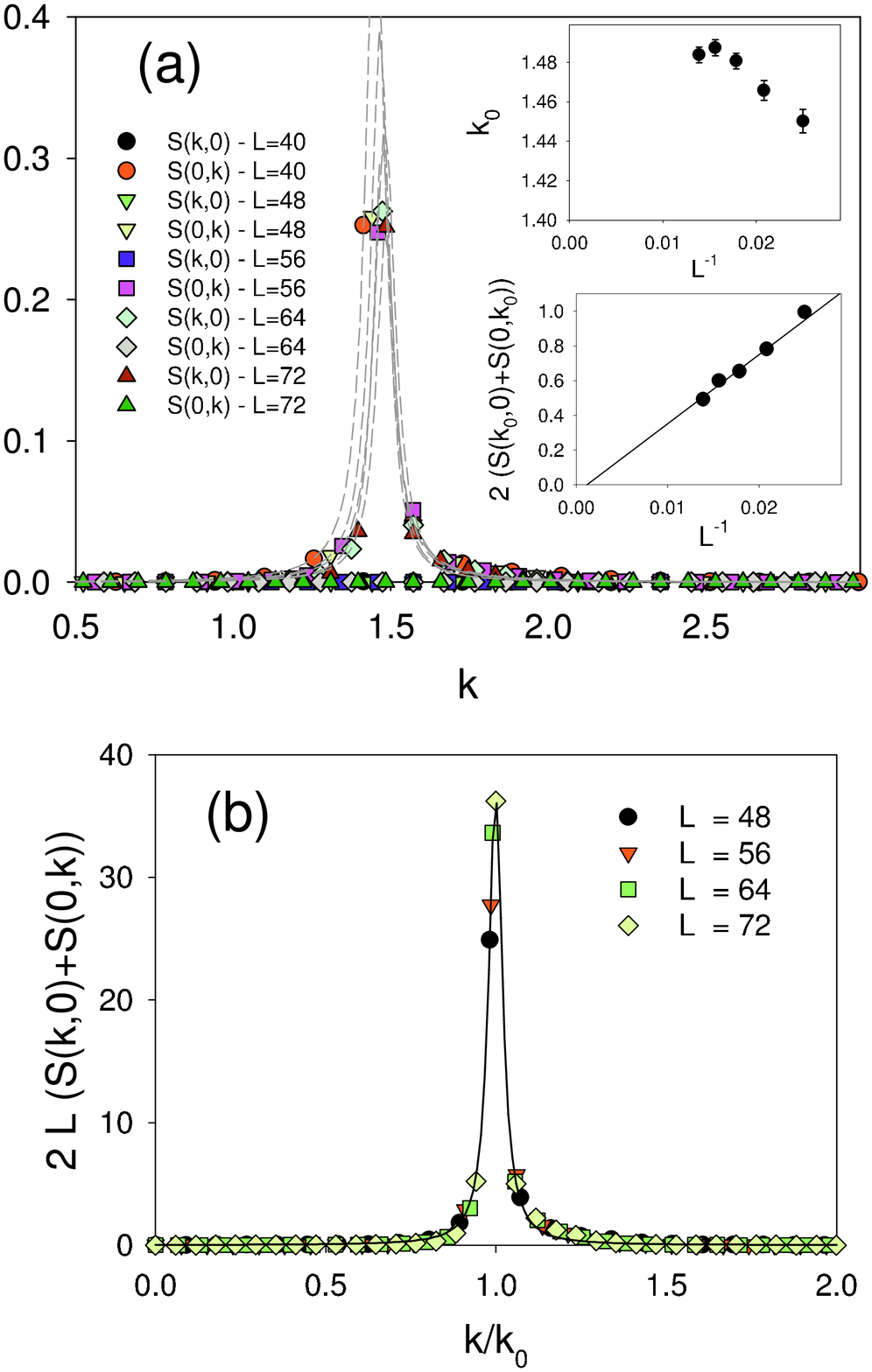}
\caption{\label{SSnematic} (a) Static structure factors $S(k,0)$
and $S(0,k)$ at $T=0.78$ for different system sizes. The dashed
lines correspond to Lorentzian fittings
$f(k)=\lambda/((k-k_0)^2+\lambda^2)$. The upper inset shows the
fitted parameter $k_0$ vs. $L^{-1}$. The lower inset shows that
the maximum of $S(\vec{k})$ scales as $L^{-1}$. (b) $2\; L
(S(k,0)+S(0,k))$  vs $k/k_0$ for the same numerical data in (a);
the continuous line is a Lorentzian fitting. The data collapse of
the different curves shows that the whole curve $S(\vec{k})$
scales as $L^{-1}$, as predicted by Eq.(\ref{Snematic}).}
\end{center}
\end{figure}

\begin{figure}
\begin{center}
\includegraphics[scale=0.32,angle=-90]{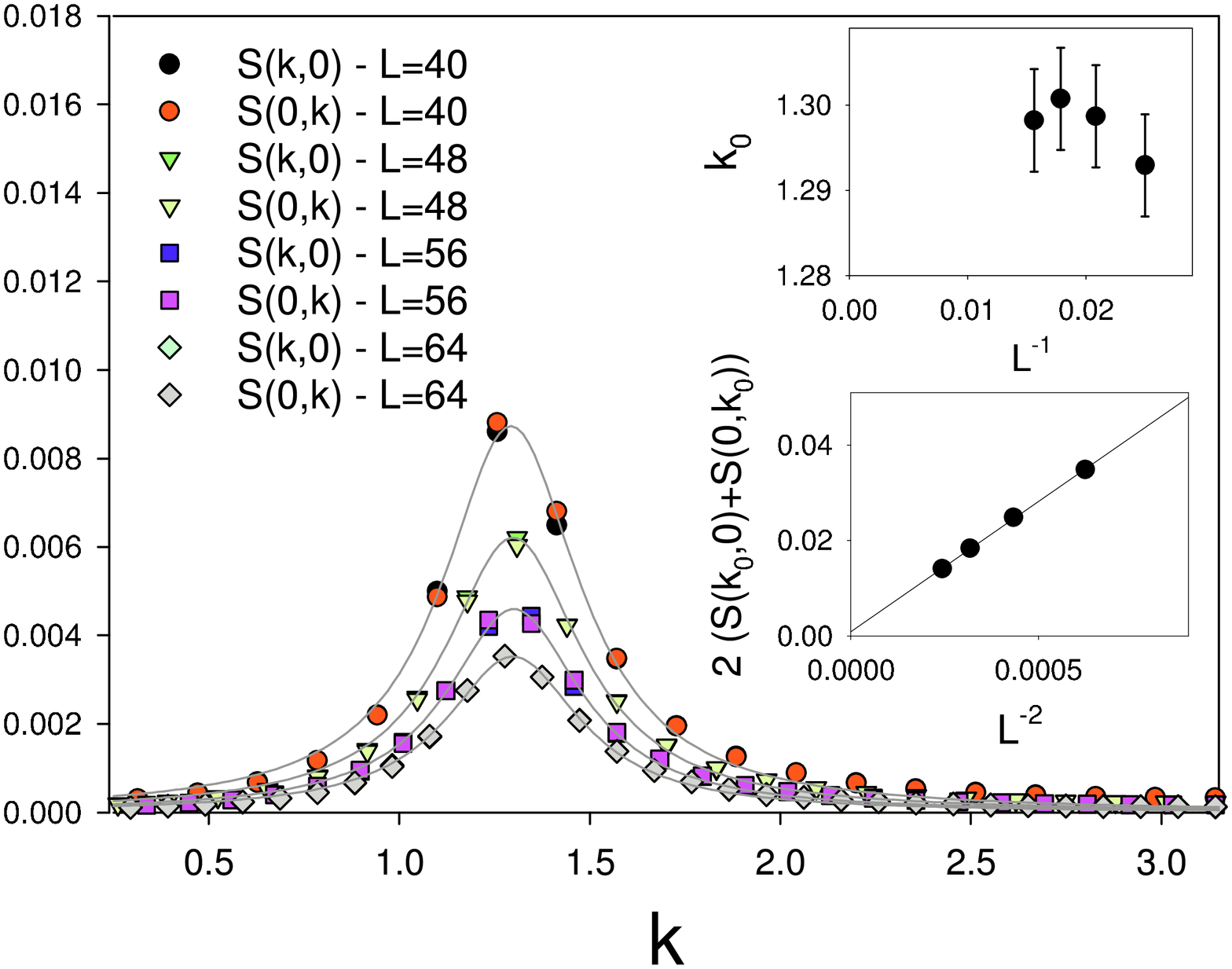}
\caption{\label{SStetragonal} Static structure factors $S(k,0)$
and $S(0,k)$ at $T=0.95$ for different system sizes. The
continuous lines correspond to Lorentzian fittings
$f(k)=\lambda/((k-k_0)^2+\lambda^2)$. The upper inset shows the
fitted parameter $k_0$ vs. $L^{-1}$. The lower inset shows that
the maximum of $S(\vec{k})$ scales as $L^{-2}$. It can be also
shown that the whole curve $S(\vec{k})$ scales as $L^{-2}$, as
predicted by Eq.(\ref{Stetragonal}).}
\end{center}
\end{figure}

We performed numerical calculations of the structure factor in the
three different phases for different system sizes up to $L=72$.
First we calculated it in the low energy phase by letting the
system to equilibrate from a vertical $h=2$ ground state
configuration at $T=0.77$, just below the estimated transition
temperature between the low and the intermediate energy phases
$T_1 \approx 0.772$ (see next section). We found that $S(\vec{k})
=c\; \delta_{k_y,0}\;  ( \delta_{k_x,\pi/2}+
\delta_{k_x,-\pi/2})$, where $c$ quickly converges to a value
$c\approx 0.48$ consistently with Eq.(\ref{Sstripes}), thus
confirming the long range stripe character of the low energy
phase. Next we let the system to equilibrate in the intermediate
phase at $T=0.78$. In Fig.\ref{SSnematic}a we plot $S(k,0)$ and
$S(0,k)$ for different system sizes. We observe that, while one of
the two functions remains almost zero for all $k$, the other
presents two symmetric peaks at $k=\pm k_0$ (only one is shown in
the figure) which are very well fitted by a Lorentz function, in
agreement with Eq.(\ref{Snematic}). We see that $k_0$ saturates
into a value $k_0\approx 1.48$ while the maxima (or equivalently
the order parameter $\eta_s$) goes to zero as $L^{-1}$. Moreover,
the data collapse of the curves for different values of $L$ when
$2\; L (S(k,0)+S(0,k))$ is plotted vs $k/k_0$ (see
Fig.\ref{SSnematic}b) shows that the whole curves scale as
$L^{-1}$ in agreement with Eq.(\ref{Snematic}); this confirms that
correlation length does not depend on $L$ in the thermodynamic
limit and that the intermediate phase does not present LRO, thus
confirming its nematic character. Finally, in
Fig.\ref{SStetragonal} we show $S(k,0)$ and $S(0,k)$ for different
system sizes after equilibration in the tetragonal liquid at
$T=0.95$. We see that both functions present two symmetric peaks
at $k=\pm k_0$ (only those at $k>0$ are shown in the figure) which
are very well fitted by  Lorentz functions and that the maxima
scale as $L^{-2}$, in excellent agreement with
Eq.(\ref{Stetragonal}). Hence, we see that the stripe order
parameter allows, not only to distinguish long range orientational
order, but also discriminate between the tetragonal and the
nematic phases through its finite size scaling.

\begin{figure}
\begin{center}
\includegraphics[scale=0.3,angle=-90]{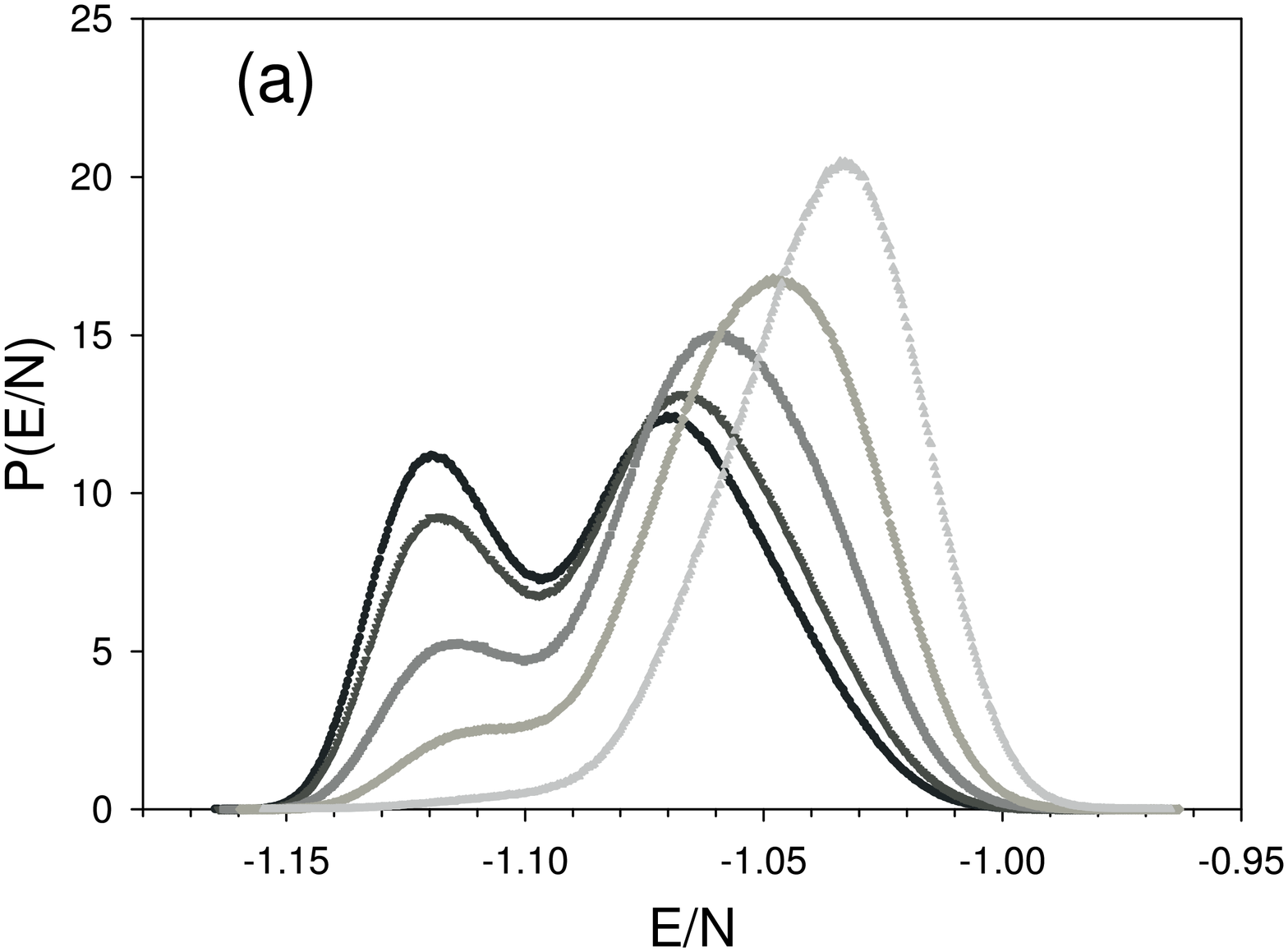}
\includegraphics[scale=0.3,angle=-90]{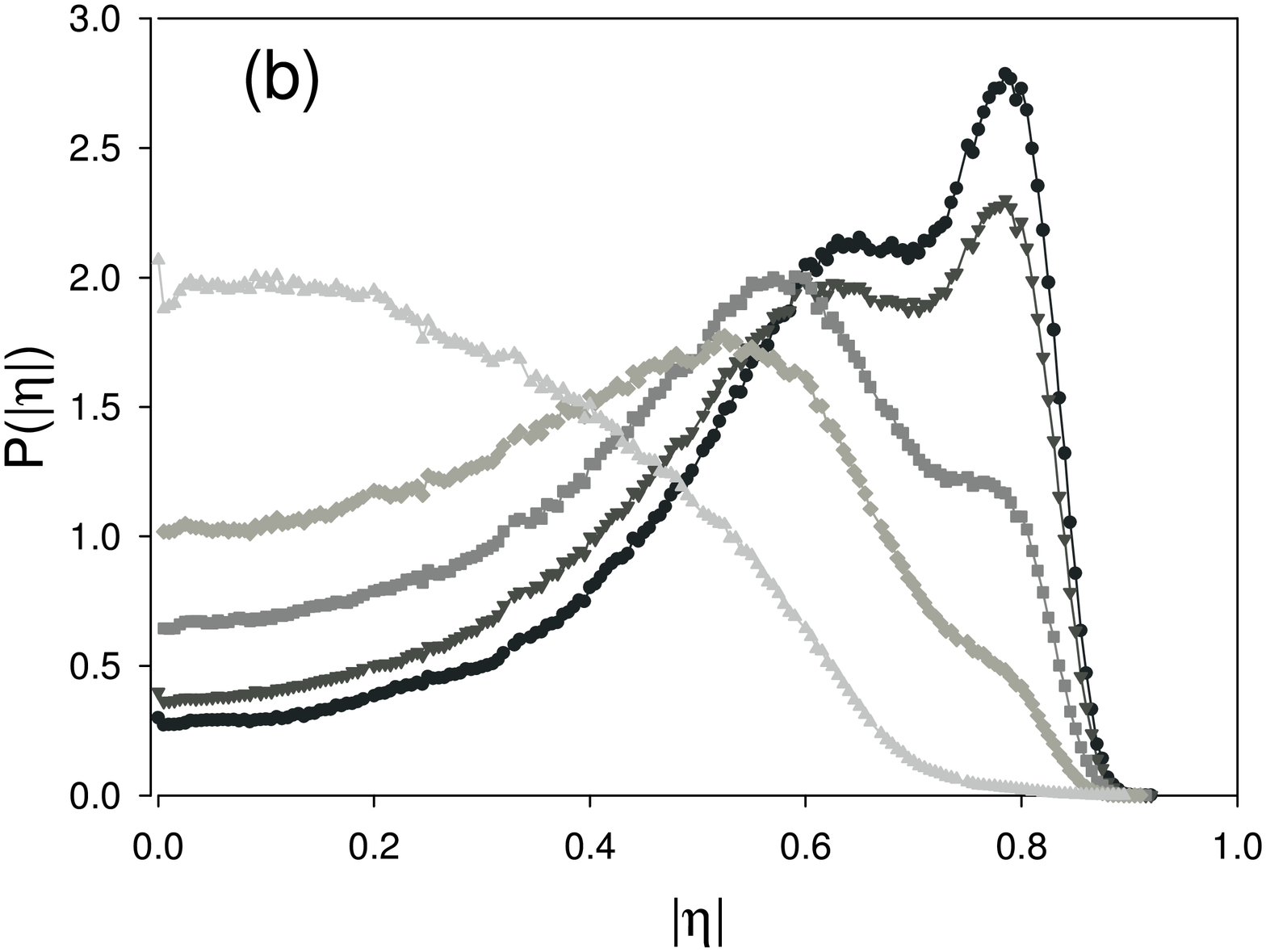}
\caption{\label{fig7} Histograms for $\delta=2$ and $L=64$, for
temperatures around the nematic-tetragonal transition. (a) Energy
per spin; temperatures from left to right: $T=0.804$, $0.805$,
$0.807$, $0.810$ and $0.815$. (b) Orientational order parameter;
same temperatures as in (a), but from right to left.}
\end{center}
\end{figure}

 It is worth mentioning that the two-peak structure in $P(E)$ and
 $P(\eta)$
associated with the nematic-tetragonal phase transition can only
be appreciated in both type of histograms only for system sizes $L
\geq 40$. For system sizes $L<40$ the two peaks are so close to
each other that the phase transition cannot longer be resolved. In
that situation the tetragonal and the nematic structures are
almost indistinguishable, the latter appearing as a slightly
elongated tetragonal one\cite{CaStTa2004}. While this strong
finite size effect disappears for system sizes $L \geq 40$, there
still remain similar finite size effects associated with the
nematic-tetragonal phase transition up to system sizes $L=64$.
This can be seen in Fig.\ref{fig7}, where we show the histograms
for the energy and the order parameter for $L=64$ and different
temperatures around that transition. If we look at the energy per
spin histogram (Fig.\ref{fig7}a) we just see that the high energy
peak becomes skewed towards the right and broadens around $T=0.81$
(compare with Fig.\ref{fig1}). However, when looking at the
orientational order parameter histogram (Fig.\ref{fig7}b) we see
that the single peak observed for smaller system sizes (see
Fig.\ref{fig2}) at that temperature range splits for $L=64$ into
two distinct peaks, one centered at $|\eta| \approx 0.8$ and the
other at $|\eta| \approx 0.6$. In Fig.\ref{fig3}b we can see an
example of a typical spin configuration in the second case. We see
that the system still exhibits nematic order, the main difference
with the configuration of Fig.\ref{fig3}a being a higher density
of domain walls perpendicular to the underlying striped structure
(disclinations), which reduce the value of $|\eta|$. These results
suggest the existence of different nematic phases separated by
first order phase transitions between them. However, to verify
this assumption much larger system sizes would be required.

\begin{figure}
\begin{center}
\includegraphics[scale=0.4]{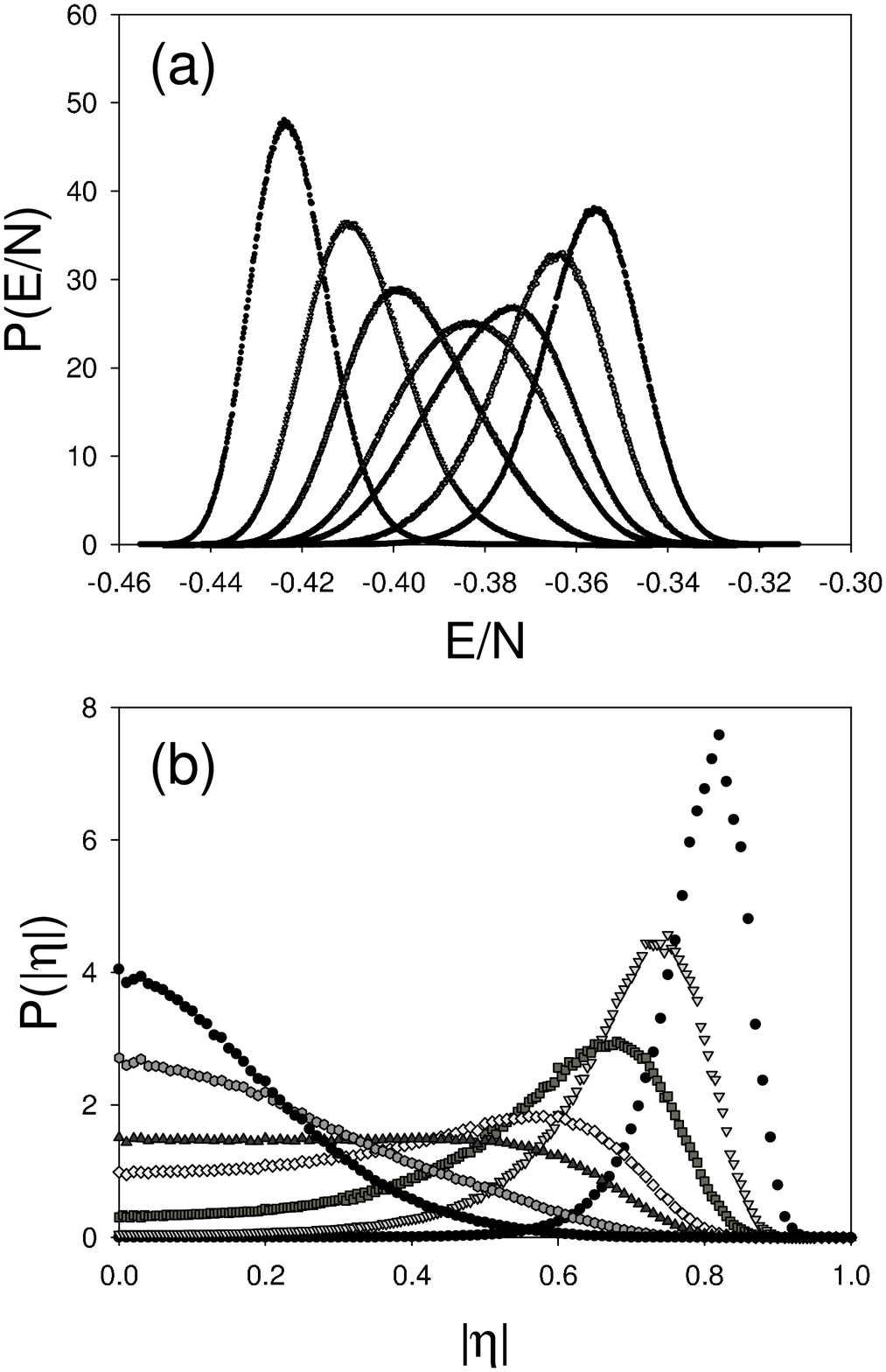}
\caption{\label{fig8} Histograms for $\delta=1$ and $L=48$.(a)
Energy per spin; temperatures from left to right: $T=0.380$,
$0.390$, $0.395$, $0.399$, $0.401$, $0.405$and $0.410$. (b)
Orientational order parameter; same temperatures as in (a), but
from right to left.}
\end{center}
\end{figure}

Next, we considered the case $\delta=1$, for which the ground
state corresponds to a striped state with $h=1$ and the transition
temperature is located around\cite{GlTaCa2002} $T=0.4$. In
Fig.\ref{fig8} we show the typical structure of the histograms for
the energy and the order parameter observed for system sizes up to
$L=48$. While we do not see in this case a double peak structure
in the energy per spin histogram (Fig.\ref{fig8}a) as in the
$\delta=2$ case, the broadening of the energy histogram around
$T=0.4$ and its skewed form at both sides of the transition,
together with a slight double peak structure of the orientational
parameter (although difficult to see, the numerical data in
Fig.\ref{fig8}b show the presence of a minimum for $T=0.401$
between $\eta=0$ and $|\eta| \approx 0.6$) suggest a single first
order phase transition. Moreover, a direct inspection of the
typical spin configuration associated with the different
distributions indicate a direct phase transition from the striped
to the tetragonal phase, without any trace of an intermediate
nematic state (notice that the location of the maximum of
$P(|\eta|)$ moves continuously towards one below the transition
point). This scenario will be confirmed by a direct
thermodynamical analysis in the next section. However, based in
the previously observed finite size effects for $\delta=2$, the
possible existence of a nematic phase in a narrow range of
temperatures for much larger system sizes cannot be excluded.

\section{Phase transitions}

We now turn our attention to the stripe-nematic and
nematic-tetragonal phase transitions in the $\delta=2$ case. To
characterize them we analyzed the finite size scaling behavior of
the moments of the energy, namely, the specific heat and the
fourth order cumulant

\begin{equation}
C(L,T) = \frac{1}{NT^2} \left( \left< H^2 \right> -\left< H
\right>^2\right)
\label{heat}
\end{equation}

\begin{equation}
V(L,T) = 1- \frac{\left< H^4 \right>}{3\left< H^2 \right>^2}
\label{cumulant}
\end{equation}

\noindent To check the simulations results we also calculated the
average energy per spin $u(T)\equiv -\left< H \right>/N$ and
compared its derivative with respect to temperature with
Eq.(\ref{heat}); both results coincide within the numerical error.
We also analyzed the average of the absolute value of the
orientational order parameter $\left<\left|\eta\right| \right>$
and the associated susceptibility

\begin{equation}
\chi_\eta(L,T) = \frac{N}{T} \left( \left< \eta^2 \right> -\left<
\left|\eta\right|\right>^2\right) \label{chi}
\end{equation}

\begin{figure}
\begin{center}
\includegraphics[scale=0.42]{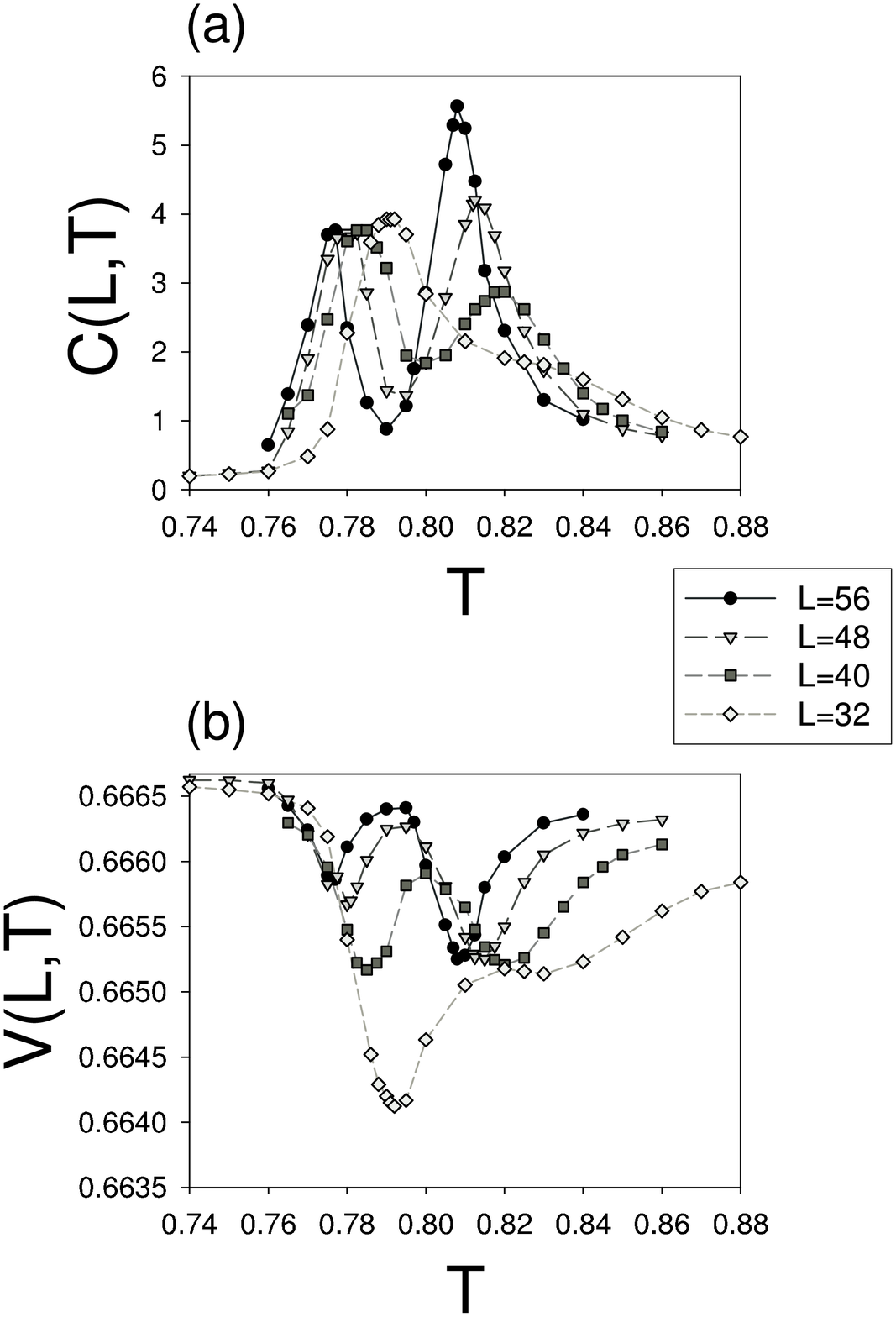}
\caption{\label{fig9} Moments of the energy for $\delta=2$ and
different system sizes . (a) Specific heat Eq.(\ref{heat}); (b)
Fourth order cumulant Eq.(\ref{cumulant}).}
\end{center}
\end{figure}

\begin{figure}
\begin{center}
\includegraphics[scale=0.42]{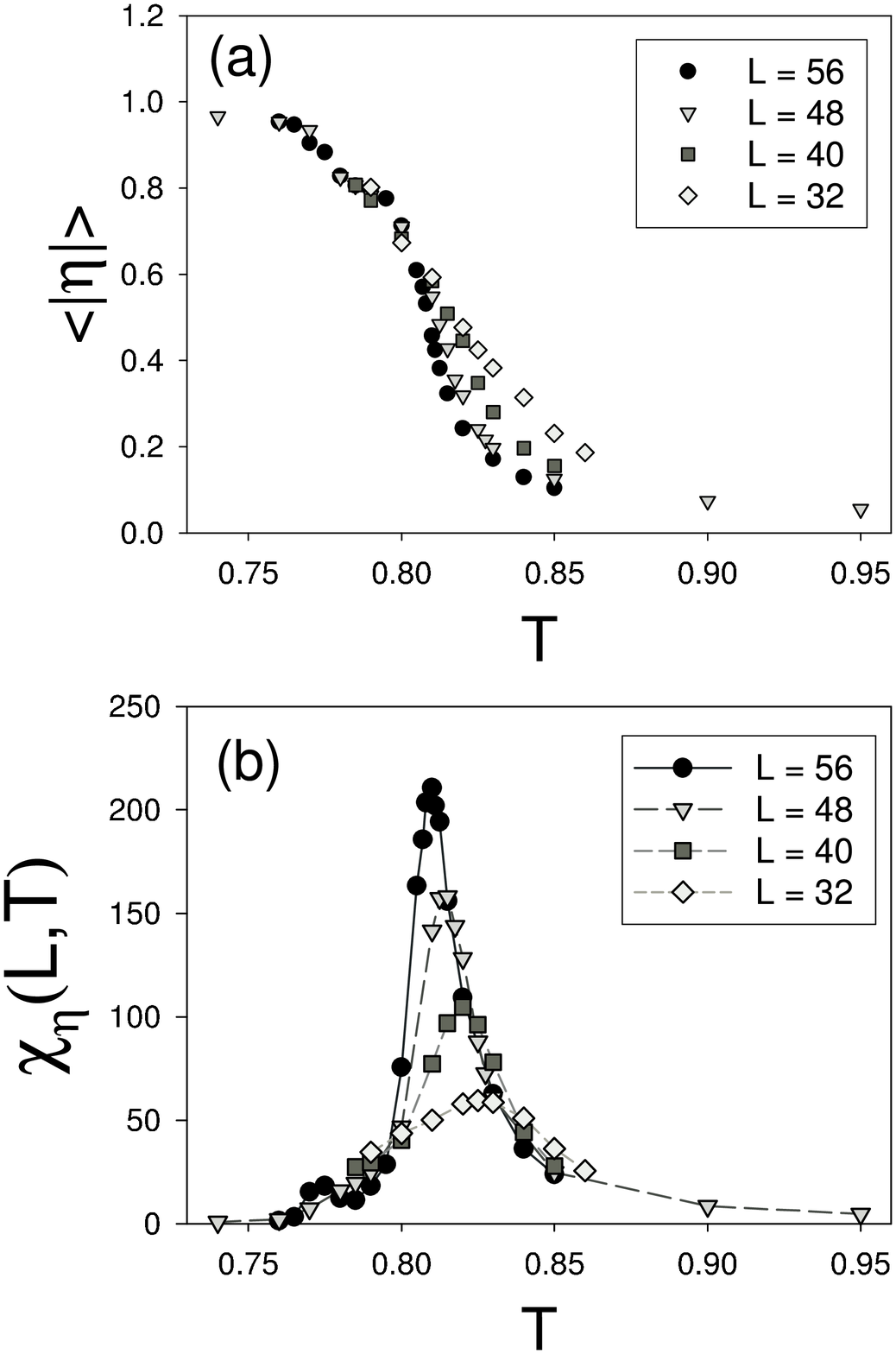}
\caption{\label{fig10} Moments of the orientational order
parameter Eq.(\ref{eta}) for $\delta=2$ and different system
sizes. (a) Average of the absolute value; (b) Associated
susceptibility Eq.(\ref{chi}).}
\end{center}
\end{figure}

All these quantities were obtained from the corresponding
histograms for system sizes up to $L=64$. In Fig.\ref{fig9} we
show the behavior of the energy moments and in Fig.\ref{fig10} of
the average order parameter and its associated susceptibility (the
results for $L=64$ are not shown for clarity).

\begin{figure}
\begin{center}
\includegraphics[scale=0.3,angle=-90]{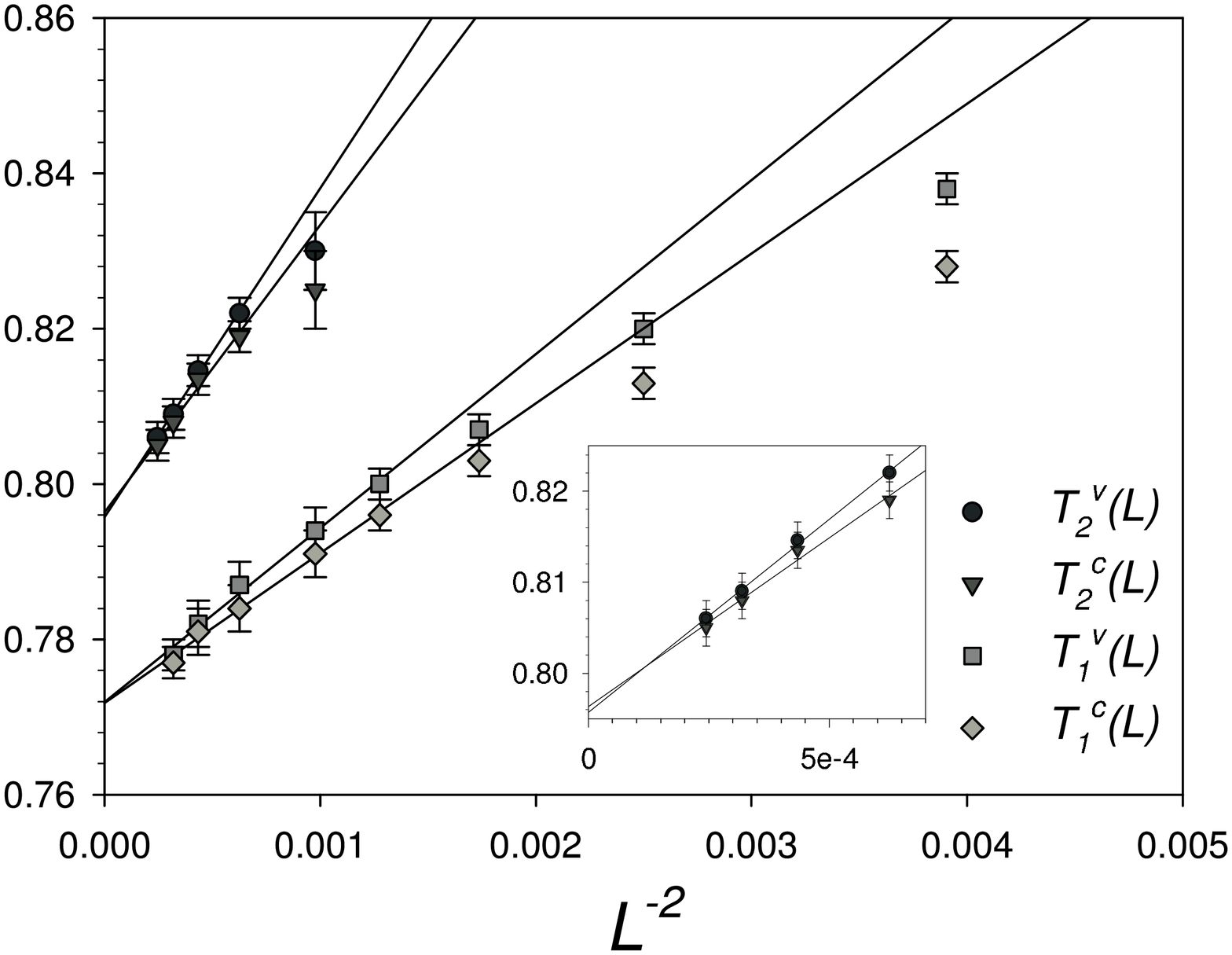}
\caption{\label{fig11} Finite size scaling of the pseudo-critical
temperatures for the maxima of the specific heat ($T_1^c$  and
$T_2^c$) and the minima of the fourth order cumulant ($T_1^v$ and
$T_2^v$) for $\delta=2$ and system sizes up to $L=64$. The inset
shows a zoom of the same curves around $T_2$.}
\end{center}
\end{figure}

\begin{figure}
\begin{center}
\includegraphics[scale=0.3,angle=-90]{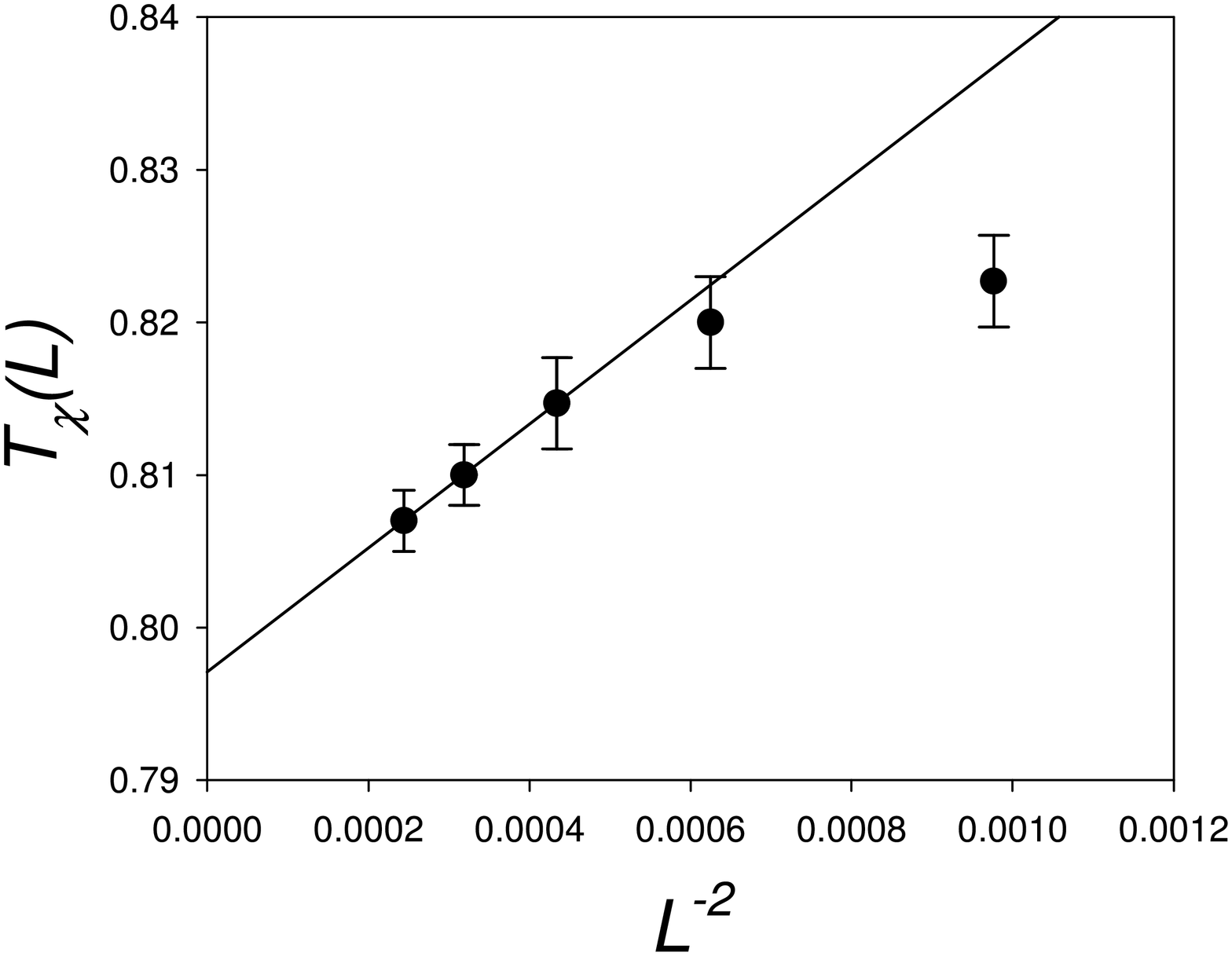}
\caption{\label{fig12} Finite size scaling of the pseudo-critical
temperatures for the maxima of the susceptibility for $\delta=2$
and system sizes up to $L=64$.}
\end{center}
\end{figure}

We see that the specific heat shows two distinctive maxima at
size-dependent pseudo--critical temperatures $T_1^c(L) <
T_2^c(L)$, while the fourth order cumulant shows two distinctive
minima at pseudo--critical temperatures $T_1^v(L) < T_2^v(L)$,
associated with the stripe-nematic and the nematic-tetragonal
transitions respectively (notice that
 both the maximum at $T_2^c$ and the minimum at $T_2^v$ are almost unperceptive for
 $L=32$). On the other hand, the orientational order parameter
 shows jumps in behavior at each transition, but the associated
 susceptibility only shows a clear size dependent maximum at a
size--dependent pseudo--critical temperature $T_2^\chi(L)$,
corresponding to the nematic--tetragonal transition. This is
consistent with the fact the $90^o$ rotational symmetry is broken
at this phase transition. Although a second peak seems to emerge
at lower temperatures for the largest size considered, it is very
small and  much more larger system sizes would be required to
asses the presence of a size dependent peak associated with the
stripe-nematic transition. From Fig.\ref{fig11} we see that the
pseudo--critical temperatures of the specific heat and the fourth
order cumulant scale as $T_1^v(L) \sim T_1 + A_1/L^2$, $T_1^c(L)
\sim T_1 + B_1/L^2$, $T_2^v(L) \sim T_2 + A_2/L^2$, $T_2^c(L) \sim
T_2 + B_2/L^2$, with $A_1 > B_1$ and $A_2 > B_2$. This is the
expected scaling behavior at first order phase transitions
\cite{ChLaBi1986,LeKo1991}. The extrapolation to
$L\rightarrow\infty$ gives the values $T_1 = 0.772 \pm 0.002$ and
$T_2=0.797 \pm 0.005$ showing that, although narrow, there is a
well resolved range of temperatures in which the nematic phase
exists in the thermodynamic limit. Moreover, from Fig.\ref{fig12}
we also see that $T_2^\chi(L) \sim T_2 + C/L^2$, where the
extrapolation to $L\rightarrow\infty$ gives the same value for
$T_2$ as the previous calculation. This behavior is also
consistent with a first order phase transition\cite{LeKo1991}.

\begin{figure}
\begin{center}
\includegraphics[scale=0.32,angle=-90]{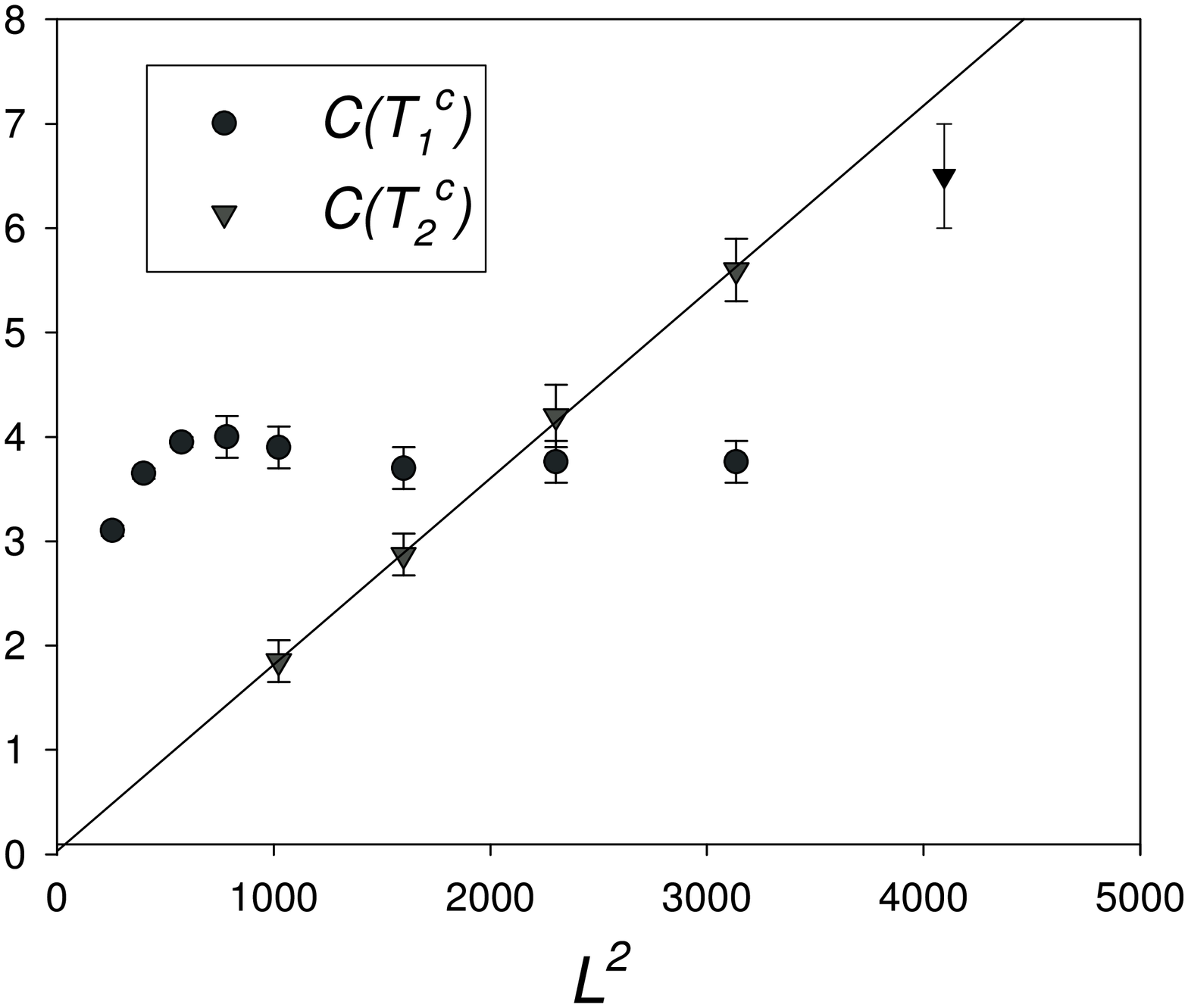}
\caption{\label{fig13} Finite size scaling of  the maxima of the
specific heat at $T_1^c$  and $T_2^c$ for $\delta=2$ and system
sizes up to $L=64$.}
\end{center}
\end{figure}

\begin{figure}
\begin{center}
\includegraphics[scale=0.32,angle=-90]{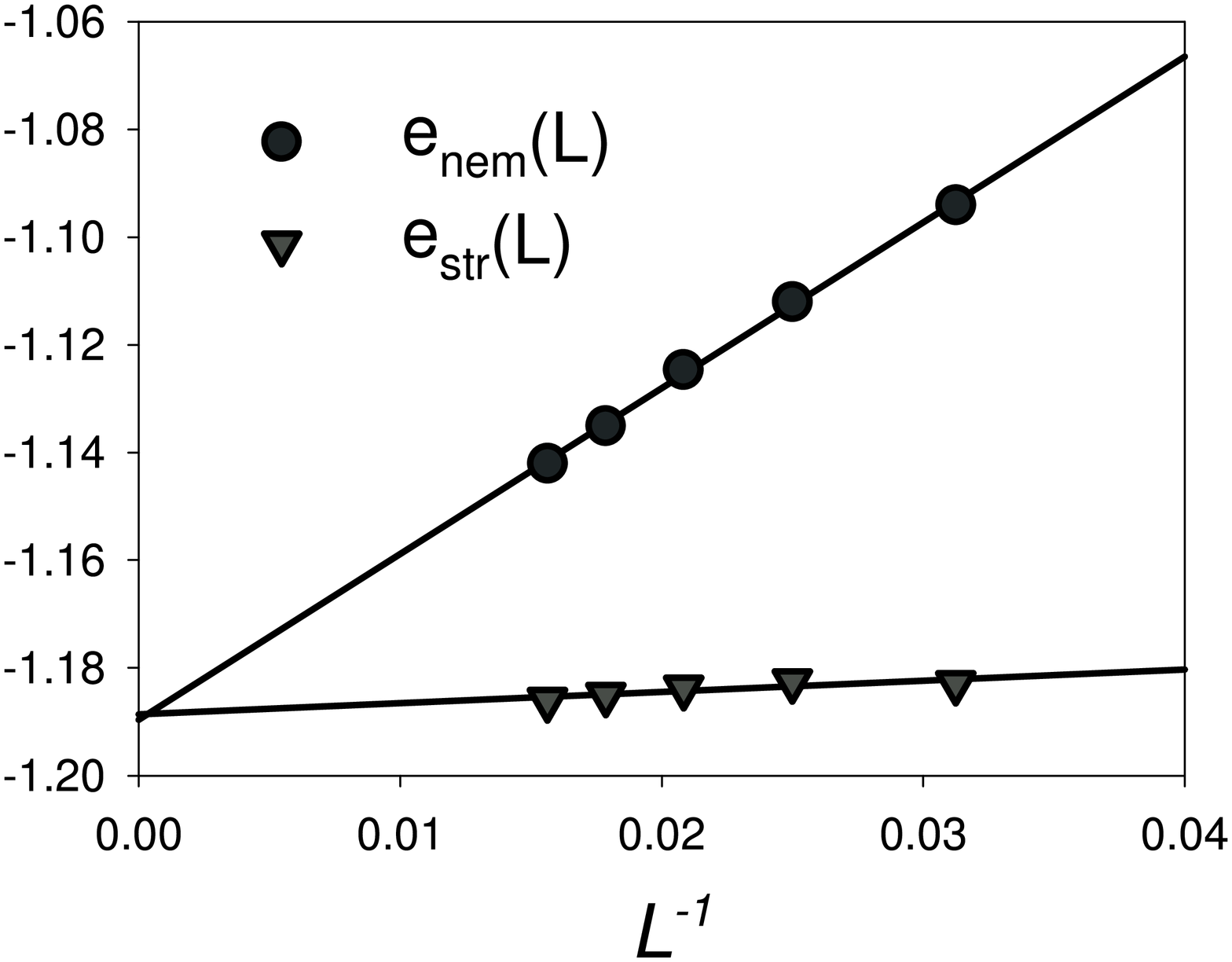}
\caption{\label{fig14} Finite size scaling of the energies per
spin at which both maxima in the corresponding probability
distribution occur, for the stripe--nematic phase transition for
$\delta=2$ and system sizes up to $L=64$; for every value of $L$
the temperature is chosen such that $P(e_{nem})=P(e_{str})$}
\end{center}
\end{figure}

However, when looking at Fig.\ref{fig9} a clear difference in
finite size scaling of the maxima of the specific heat and the
minima of the fourth order cumulant between both transitions
appears: while the minimum of $V$ at $T_2^v$ rapidly saturates at
a constant value when $L\rightarrow\infty$, the minimum of $V$ at
$T_1^v$ shows scaling behavior, approaching systematically to
$2/3$. From Fig.\ref{fig13} we see that the maximum of the
specific heat at $T_2^c$ increases as $L^2$ for system sizes up to
$L=56$, as expected in a first order transition, but it shows a
little departure from this behavior for $L=64$. However, such
departure is probably due to systematic errors introduced by the
presence of two phase transitions (see section \ref{histos}) that
cannot be resolved for small system sizes. The scaling of
$\chi_\eta(T_2^\chi)$ is similar to that of $C(T_2^2)$. On the
other hand, the maximum of $C$ at $T_1^c$ saturates clearly in a
finite value when the system size increases. It can also be shown
that $2/3 - V(T_1^v) \sim L^{-2}$. This behavior can be understood
if we look at the energy per spin histogram at the temperature
where both maxima have the same height. If we call $e_{str}(L)$
and $e_{nem}(L)$ the energies per spin at which those maxima
occur, in a first order phase transition it is expected that they
will scale as\cite{LeKo1991} $e_{str}(L) - u_{str} \sim L^{-1} $
and $e_{nem}(L) - u_{nem} \sim L^{-1} $, $u_{str}$ and $ u_{nem}$
being the specific internal energies of both phases (the striped
and the nematic ones, in this case) in the thermodynamic limit.
Then, the maximum of the specific heat is expected to scale in a
two-dimensional system as\cite{LeKo1991} $C(T_1^c) \sim L^2\,
\left( e_{nem}(L)-e_{str}(L)\right)^2$, while $2/3 - V(T_1^v)\sim
\left( e_{nem}(L)-e_{str}(L)\right)^2 + {\cal O}(L^{-2})$. In
Fig.\ref{fig14} we see that $e_{str}(L)$ and $e_{nem}(L)$ show the
expected scaling, but they converge {\it to the same value}, so
that $e_{nem}(L)-e_{str}(L) \sim L^{-1}$ and this explains the
observed behaviors of the maximum of $C$ and the minimum of $V$.
This shows that the internal energy at the stripe-nematic
transition becomes continuous in the thermodynamic limit. However,
a thermodynamical singularity still develops in that limit. This
can be seen by looking at the scaling of the free energy barrier
between both phases, which can be calculated as\cite{LeKo1991}

\begin{figure}
\begin{center}
\includegraphics[scale=0.32,angle=-90]{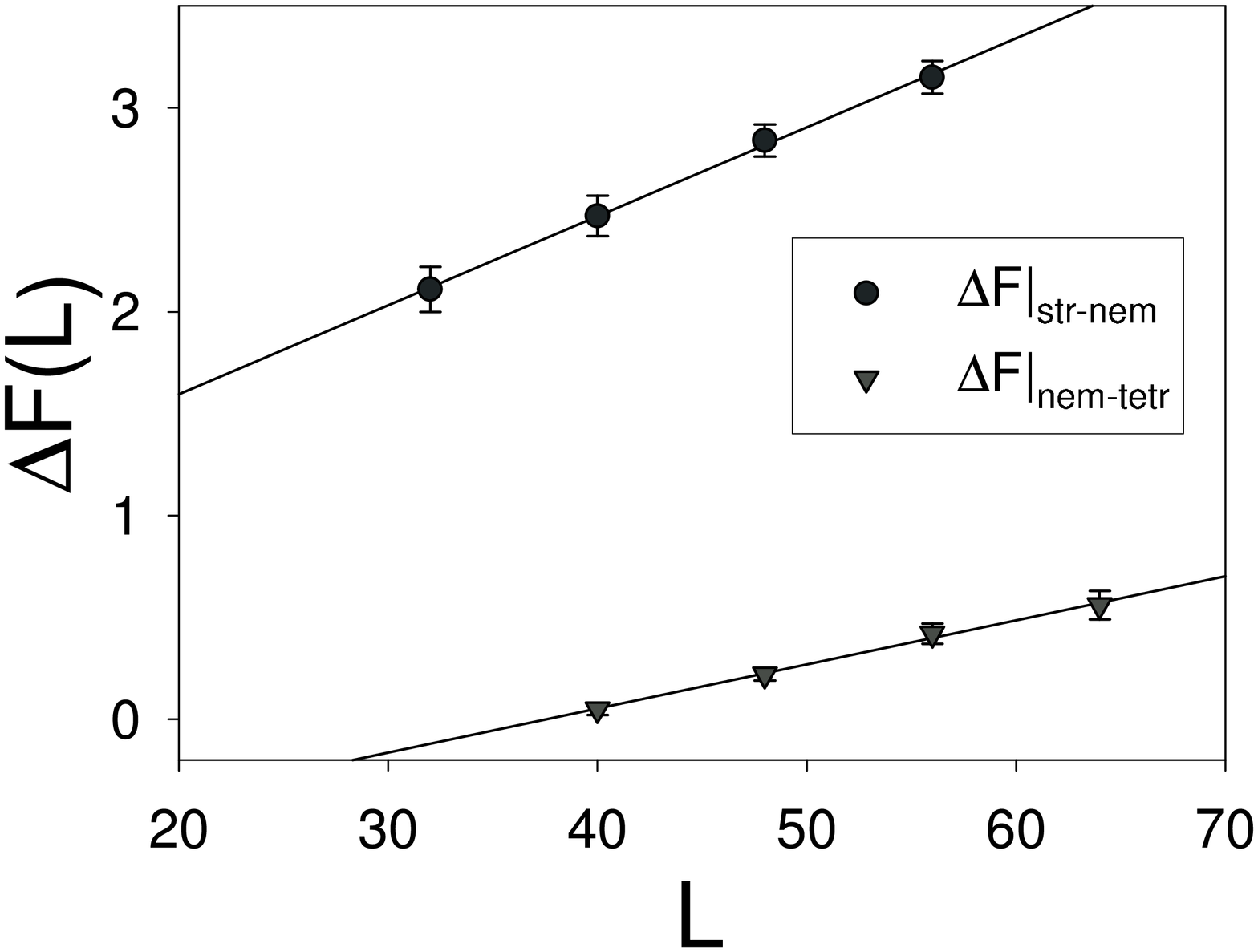}
\caption{\label{fig15} Finite size scaling of the free energy
barrier Eq.(\ref{DeltaF}) between  phases in the stripe--nematic
and the nematic--tetragonal phase transitions for $\delta=2$.}
\end{center}
\end{figure}

\begin{equation}
\Delta F = \ln{ \left( \frac{P_{max}}{P_{min}} \right)}
\label{DeltaF}
\end{equation}

\noindent where $P_{max}\equiv P(e_{nem})=P(e_{str})$ is the value
of the common maximum of energy per spin histogram and $P_{min}$
is the value of the minimum between them. In Fig.\ref{fig15} we
see that $\Delta F \sim L$ in both phase transitions, which is the
expected scaling in a first order phase transition\cite{LeKo1991}.
In particular, in the case of the stripe-nematic transition it
means that, even when both states acquire the same energy, there
is a divergent free energy barrier between them. Finally, we
analyzed the difference between the values of the parameter $\eta$
at which the maxima in the corresponding histogram occur in the
stripe-nematic phase transition $\Delta \eta (L) = \eta_{str}
-\eta_{nem}$. A finite size scaling analysis similar  to the
previous ones shows that $\Delta \eta (L) \sim \Delta \eta
(\infty) + D/L^2$, with $ \Delta \eta (\infty) =0.075 \pm 0.01$.
Hence, a finite jump in this parameter, even small, persists in
the thermodynamic limit, consistently with the discontinuous jump
in the stripe order parameter already observed in the previous
section.

A specific heat peak saturation has been observed numerically in
different systems exhibiting a KT type phase transition, such as
the 2D XY model\cite{GuBa1992} and generalizations and the 1D
Ising model with $1/r^2$ ferromagnetic
interactions\cite{LuMe2001,ImNe1988}. This last model is
particularly interesting, since it presents long range order in
the low temperature phase (at variance with the XY models, which
present only short range order) and  the order parameter
(magnetization) is discontinuous at the
transition\cite{AiChChNe1988}, in an analogous way to the model we
are considering here. However, an important difference should be
noted: in the above mentioned examples of the KT transition the
extrapolation of the pseudo critical temperature at which the
maximum of the specific heat is located appears slightly above the
critical temperature.

Next we considered the tetragonal-stripe phase transition in the
case $\delta=1$. As we have seen in section \ref{histos}, the
equilibrium histograms suggest a weak first order phase
transition. Since a finite size scaling approach would require
system sizes much larger than the allowed by the present
computational capabilities, to confirm the supposed first order
nature of the transition we considered another type of approach.

\begin{figure}
\begin{center}
\includegraphics[scale=0.45,angle=0]{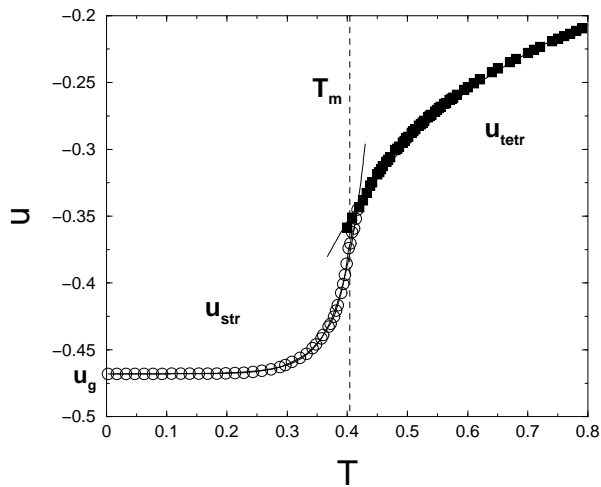}
\caption{\label{fig_energiainterna} Internal energy per spin $u$
versus $T$ for $\delta=1$ and $N=48\times48$. Also shown best fits
to the low temperature stripe and high temperature tetragonal
regions, as described in the text. The best fit parameters for
$u_{str}$ are $a_1=1952.54; a_2=10.8491;a_3=10.8441;a_4=5.73016;
a_5=-58.0638$ and $ a_6=7.34061$, and for $u_{tetr}$ are
$b_0=-1.47926;   b_1=0.0817682; b_2=0.0559538 $ and
$b_3=-3.05417$. The ground state energy is $u_g=-0.4677$.}
\end{center}
\end{figure}

\begin{figure}
\begin{center}
\includegraphics[scale=0.45,angle=0]{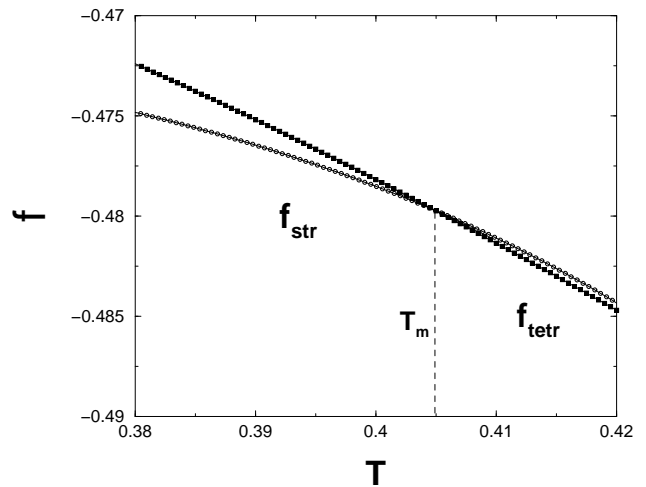}
\caption{\label{fig_energialivre}Free energy per spin $f$ versus
$T$ for $\delta=1$. The stripe and tetragonal free energies cross
each other around $T_{m}=0.404$. The functions obtained from
equation (\ref{energia_livre_spin}) are
$f_{str}=u_g+\frac{a_1}{1-a_2}T^{a_2}+\frac{a_3}{1-a_4}T^{a_4}+\frac{a_5}{1-a_6}T^{a_6}$
and
$f_{tetr}=\frac{b_0}{b_1}T\textrm{ln}(\textrm{cosh}\frac{b_1}{T})+\frac{b_2}{b_3}T\textrm{ln}(\textrm{cosh}\frac{b_3}{T})-T\textrm{ln}(2)$,
where $\beta_0=0$ for the tetragonal and $\beta_0=\infty$ for the
striped phase.}

\end{center}
\end{figure}

\begin{figure}
\begin{center}
\includegraphics[scale=0.45,angle=0]{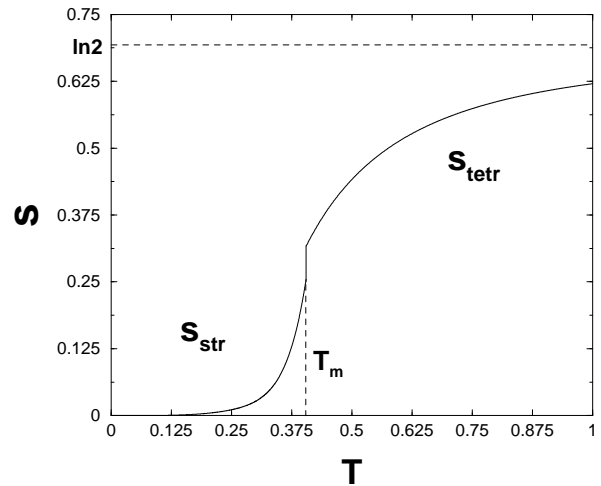}
\caption{\label{fig_entropia} Entropy per spin $s$ versus $T$ for
$\delta=1$. The slight discontinuity at the transition temperature
is consequence of a weak first order phase transition.}
\end{center}
\end{figure}

Since we know the value of the ground state energy $u_g$ and also
the asymptotic infinite temperature internal energy, we can obtain
the free energy for different temperatures integrating the
measured internal energies. We checked that the internal energy is
independent of system size for $48 \le L \le 64$, and we chose a
$N=48 \times 48$ lattice in order to measure internal energy. The
equilibrium internal energy as a function of temperature for the
stripe and tetragonal phases are shown in
Fig.(\ref{fig_energiainterna}). Using this energy curve we fit the
stripe energy  by a polynomial function
$u_{str}(T)=u_g+a_1T^{a_2}+a_3T^{a_4}+a_5T^{a_6}$ and the
tetragonal energy with a sum of hyperbolic functions
$u_{tetr}(T)=b_0\textrm{tanh}(\frac{b_1}{T})+b_2\textrm{tanh}(\frac{b_3}{T})$,
whose parameters are shown in the caption. By means of the
thermodynamical relation:

\begin{eqnarray}
\beta f(\beta)= \beta_0f_0(\beta_0) +
\int_{\beta_0}^{\beta}u(\beta)d\beta, \label{energia_livre_spin}
\end{eqnarray}

where $\beta =1/T$ and $\beta_0=1/T_0$, we obtained the free
energy per particle, $f(\beta)$ for the striped and tetragonal
phases (shown in Fig. (\ref{fig_energialivre})). The transition
temperature is obtained by imposing $f_{str}(T)=f_{tetr}(T)$,
which gives $T_m=0.404$. The entropy is obtained directly by the
thermodynamical relation $s=-\frac{df}{dT}$ and the result is
shown in the Fig.(\ref{fig_entropia}). A weak first order
transition is apparent, in agreement with the behavior of the
energy and order parameter histograms of Fig.\ref{fig8}.

These results for the thermodynamic functions point to a unique
phase transition for $\delta=1$, from a high temperature
tetragonal liquid phase to a low temperature striped phase. There
is no trace of a third Ising nematic phase as for $\delta=2$, at
least for $48 \le L \le 64$ system sizes.

This conclusion is natural in this case where the equilibrium low
temperature phase corresponds to stripes of width $h=1$. Once
defects begin to emerge due to thermal fluctuations, disclinations
appear naturally and the orientational order will rapidly decay ,
at variance with the expected behavior for wider stripes which
will be more stable to the appearance of topological defects.

\section{Metastability}

The presence of diverging free energy barriers between the
different phases found in the previous section for $\delta=2$ lead
us to consider the possible existence of dynamical metastable
(i.e., quasi--stationary) states. To this end, we performed
cooling and heating numerical calculations according to the
following protocol. We first let the system equilibrate at a
temperature high enough to ensure that it is in the tetragonal
liquid phase. We then cooled the system at a fixed cooling rate
$r$, that is, the temperature was decreased as $T(t) = T_0 -r \;
t$, where $T_0$ is the initial temperature and the time $t$ is
measured in MCS. The temperature was reduced in every MC run down
to a value well below the range associated with the different
phase transitions, while recording at different points the energy
and the orientational order parameter $|\eta|$ of the system.
Every curve was then averaged over different initial conditions
and different sequences of the random noise. For every system size
we calculated those curves for decreasing cooling rates until they
became independent of the cooling rate. In this way we simulated a
process of quasi-static cooling. Once we obtained the quasi-static
cooling curves we performed a quasi-static heating from zero
temperature  up to high temperatures, starting from the ground
state and using the same protocol and the same rate $r$.

\begin{figure}
\begin{center}
\includegraphics[scale=0.48]{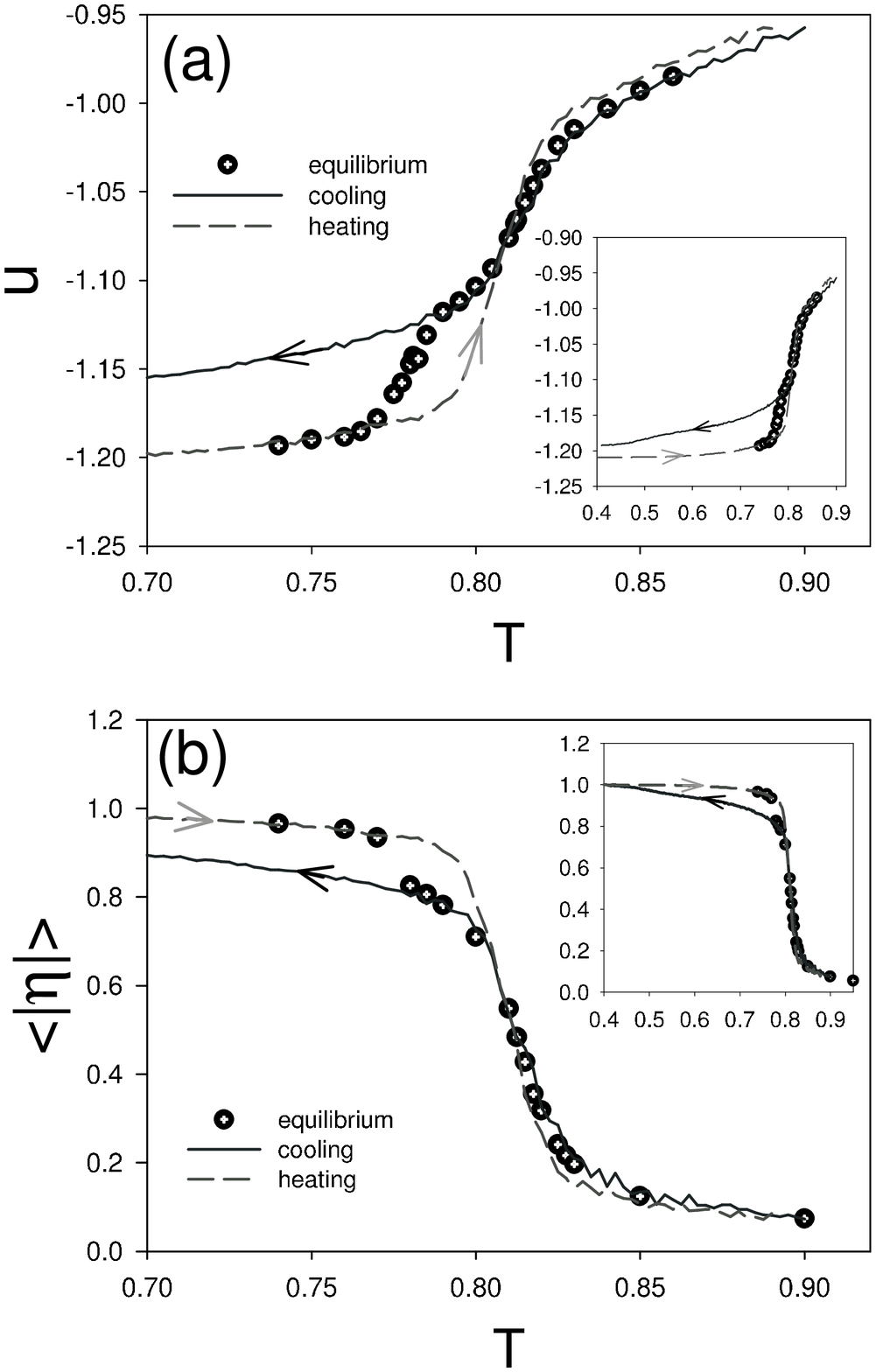}
\caption{\label{figcooldelta2} Cooling-heating curves for
$\delta=2$, $L=48$ and $r=10^{-6}$. The insets show the same
curves for a wider range of temperatures. (a) Average energy per
spin (b) Average of the absolute value of the orientational order
parameter.}
\end{center}
\end{figure}

In fig.\ref{figcooldelta2} we see an example of the quasi-static
cooling-heating curves obtained for $L=48$ with a cooling rate $r=
10^{-6}$; the results are compared with the equilibrium curves
obtained in the previous section. Similar results were obtained
for system sizes up to $L=64$. We do not observe super cooled
tetragonal liquid states below $T_2$, but this is probably due to
the relatively small values of the associated free energy barriers
for the system sizes here considered. On the other hand, we
observe a strong metastability associated with the stripe--nematic
phase transition, which is consistent with the observed large free
energy barriers. In particular, we see that the supercooled
nematic state is observed down to temperatures well below $T_1$.
To verify the quasi-equilibrium nature of the metastable nematic
states, we calculated the two-times autocorrelation function

\begin{equation}
C(t_1,t_2) \equiv \frac{1}{N} \sum_i \; \left< S_i(t_1) \,
S_i(t_2) \right>
\end{equation}

\noindent after a quasi--static cooling to different temperatures
$T<T_1$, down to $T=0.5$, for different pairs of values $t_1<t_2$.
In all the cases we found that $C(t_1,t_2)$ depends only on the
difference $t_2-t_1$ (as expected for a quasi--stationary
process), for time scales up to $t= 10^6$ MCS. For larger time
scales the nematic phase finally decay into the striped one,
apparently through a nucleation process. These results are not
shown here, but a complete description of those calculations will
be published shortly\cite{CaStTa2005b}.

\section{Discussion}

The main contribution of this paper are the
 several evidences of the existence of nematic
phases in an ultrathin magnetic film model, at least for some
range of values of the exchange to dipolar intensities ratio.
These results are in agreement with one of the two possible
scenarios of the critical behavior of those systems predicted by
the continuum approximation of Abanov and
co--workers\cite{AbKaPoSa1995}. Let us discuss first the
stripe-nematic phase transition. Although the finite size scaling
behavior of different quantities around that transition agree with
that expected in a first order transition, both the fact that the
energy becomes continuous and the saturation of the specific heat
maximum in the thermodynamic limit are unusual in a simple first
order transition, suggesting a more complex phenomenology. In
fact, those properties strongly resemble those observed in the
$1/r^2$ one dimensional Ising model transition point, namely,
continuous energy, discontinuous order parameter and saturation of
the specific heat maximum. These analogies suggest that some KT
type mechanism, probably  associated with the unbinding of
dislocations, could be responsible of the observed phenomenology,
in agreement with Abanov and coworkers prediction. The observed
first-order-like finite size scaling behavior appears to be
related with the discontinuous change of the orientational order
parameter at the transition point in the thermodynamic limit,
which suggests the existence of a finite density of dislocation
pairs. Indeed, we observed the presence of an increasing number of
dislocations pairs (bridges) in the striped phase at $T=T_1$ but,
up to $L=72$ system sizes we didn't find any evidence of QLRO
(algebraic decaying correlations).  One important consequence  of
such first-order-like phenomenology is the existence of diverging
free energy barriers at that point, which has an important
physical implication, namely, the possibility to obtain highly
stable supercooled nematic states through cooling. This opens the
possibility of having complex slow dynamical behaviors after a
sudden quench, such as those observed in supercooled molecular
liquids. We are investigating this problems and the results will
be presented in a forthcoming publication\cite{CaStTa2005b}.

Concerning the nematic-tetragonal phase transition, our results
suggest a much more complex behavior than the expected from Abanov
et al. results, who conjectured a second order phase transition.
Although our results are inconclusive due to the presence of
strong finite size effects, they suggest a complex phase diagram
with multiple nematic phases, separated by first order transition
lines, similar to that encountered by Grousson et
al\cite{GrTaVi2001} in a related model, the 3D Coulomb frustrated
Ising ferromagnet. However, the departure from the expected first
order finite size scaling observed for $L=64$ in the maximum of
the specific heat (see Fig.\ref{fig13}) and the susceptibility,
could be an indication of a behavior similar to that observed in
the stripe-nematic transition. Hence, the possibility of a
nematic-tetragonal KT phase transition,  associated with a
disclination unbinding mechanism, cannot be excluded. A similar
scenario appears in a related problem, namely, the two-dimensional
melting\cite{St1988}. On the other hand, the Hartree approximation
of the Landau-Ginzburg version of the present model predicts a
fluctuation induced first order phase transition\cite{CaStTa2004}
for any value of $\delta$.

The case of small exchange/dipolar ratio is simpler in the sense
that only one transition is observed from a stripe phase directly
to a tetragonal liquid phase. We obtained evidence that this
transition is a weak first order one.

Although we have analyzed the critical behavior of this system for
relatively small values of the exchange/dipolar ratio ($\delta=1$
and $\delta=2$), the present results may help to understand the
critical behavior for larger values. Booth and
coworkers\cite{BoMaWhDe1995} analyzed the behavior of the specific
heat and the orientational order parameter for $\delta=3$ and
$\delta=4.45$ and system sizes up to $L=64$. While their results
suggest a continuous stripe-tetragonal phase transition, numerical
results based on time series analysis from Casartelli and
coworkers\cite{CaDaRaRe2004}, for the same parameters and system
sizes, indicate that the transition is first order. For those
values of $\delta$ the stable low temperature phase is composed by
stripes with a larger width ($h=4$), so much more larger system
sizes would be required to allow the appearance of a large density
of topological defects (dislocations and disclinations of the
stripes). Therefore, for larger system sizes one could expect a
scenario similar to that observed for $\delta=2$, namely, the
appearance of nematic order in a narrow range of temperatures.

The experimental verification of the different transitions and
phases suggested by theory and simulations is an important
challenge which can be reached in the near future due to the
emergence of new observation techniques.

 Fruitful discussions with
A. Cavagna and T. Grigera are acknowledged. We thank very useful
suggestions and comments from anonymous referees. This work was
partially supported by grants from CONICET (Argentina), Agencia
C\'ordoba Ciencia (Argentina), SeCyT, Universidad Nacional de
C\'ordoba (Argentina), CNPq (Brazil) and ICTP grant NET-61
(Italy).


\end{document}